\documentclass[11pt]{article}
\usepackage{amsmath, amssymb}
\usepackage{graphicx}
\usepackage{geometry}
\usepackage[numbers]{natbib}
\bibpunct{[}{]}{,}{n}{}{;}
\usepackage{times}
\usepackage[T1]{fontenc}
\usepackage{mathptmx}
\usepackage[hidelinks]{hyperref}

\begin{document}
\title{Interface-Bound States and Majorana Zero Modes in Lateral Heterostructures of Bi$_2$Se$_3$ and Sb$_2$Te$_3$ with Proximity-Induced Superconductivity}
\author{Yoonkang Kim\thanks{E-mail: yoonkkim04@snu.ac.kr}}
\date{}
\maketitle

\begin{center}
Department of Physics and Astronomy, Seoul National University,\\
1, Gwanak-ro, Gwanak-gu, Seoul, 08826, South Korea
\end{center}

\begin{abstract}
We present a comprehensive investigation into the emergence of interface-bound states, particularly Majorana zero modes (MZMs), in a lateral heterostructure composed of two three-dimensional topological insulators (TIs), Bi$_2$Se$_3$ and Sb$_2$Te$_3$, under the influence of proximity-induced superconductivity from niobium (Nb) contacts. We develop an advanced two-dimensional Dirac model for the topological surface states (TSS), incorporating spatially varying chemical potentials and s-wave superconducting pairing. Using the Bogoliubov-de Gennes (BdG) formalism, we derive analytical solutions for the bound states and compute the local density of states (LDOS) at the interface, revealing zero-energy modes characteristic of MZMs. The topological nature of these states is rigorously analyzed through winding numbers and Pfaffian invariants, and their robustness is explored under various physical perturbations, including gating effects. Our findings highlight the potential of this heterostructure as a platform for topological quantum computing, with detailed predictions for experimental signatures via tunneling spectroscopy.
\end{abstract}

\section{Introduction}
Topological insulators (TIs) have emerged as a cornerstone of modern condensed matter physics, characterized by insulating bulk states and gapless, spin-momentum-locked surface states protected by time-reversal symmetry \cite{Hasan2010, Qi2011}. Among the most studied TIs are Bi$_2$Se$_3$ and Sb$_2$Te$_3$, which exhibit Dirac-like topological surface states (TSS) but differ in their intrinsic doping: Bi$_2$Se$_3$ is naturally n-type, while Sb$_2$Te$_3$ is p-type \cite{Zhang2009, Chen2010}. This doping disparity provides a unique opportunity to engineer lateral heterostructures where the interface between these materials can host exotic quantum states.

When superconductivity is induced in the TSS via proximity to a conventional s-wave superconductor like niobium (Nb), the system can realize a topological superconductor, potentially supporting Majorana zero modes (MZMs) \cite{Fu2008}. MZMs are zero-energy quasiparticles with non-Abelian statistics, making them prime candidates for fault-tolerant quantum computing \cite{Kitaev2003, Alicea2012}. In this work, we focus on a lateral heterostructure of Bi$_2$Se$_3$ and Sb$_2$Te$_3$ with Nb-induced superconductivity, aiming to analytically investigate the conditions under which MZMs emerge at the interface and how they can be tuned via external parameters such as gating.

Our approach combines advanced analytical methods, including the Dirac-BdG formalism, field-theoretic techniques, and topological invariant calculations, to provide a rigorous and detailed analysis of the interface-bound states. We also explore the effects of gating on the chemical potential profile, offering a pathway to experimentally control the topological phase transitions. The results are contextualized with potential experimental signatures, particularly zero-bias peaks (ZBPs) in tunneling spectroscopy, which serve as hallmarks of MZMs.

\section{Analytical Framework and Results}

\subsection{Device Scheme and Physical Configuration}
The system under consideration is a lateral heterostructure consisting of Bi$_2$Se$_3$ (for $x < 0$) and Sb$_2$Te$_3$ (for $x > 0$), with the interface located at $x = 0$. Both TIs are thin films, approximately 15 quintuple layers (QL) thick, epitaxially grown on a hexagonal boron nitride (hBN) substrate to ensure high-quality interfaces. Superconductivity is induced in the TSS through proximity to two Nb pads placed on top of the Bi$_2$Se$_3$ and Sb$_2$Te$_3$ regions, respectively, without covering the interface. This setup allows the interface to remain unaffected by direct contact with the superconductor, preserving its intrinsic properties while still benefiting from the proximity effect.

For experimental probing, two gold (Au) electrodes are incorporated solely for tunneling spectroscopy:
- **Top Au Electrode**: Positioned at the Bi$_2$Se$_3$-Sb$_2$Te$_3$ interface, this electrode probes the local density of states (LDOS) through differential conductance ($dI/dV$) measurements.
- **Bottom Au Electrode**: Situated beneath the hBN substrate, this serves as the counter electrode for vertical tunneling spectroscopy, enabling a complete tunneling circuit with the Top Au.

These Au electrodes are dedicated to spectroscopy and do not contribute to gating. Instead, gating of the Bi$_2$Se$_3$ and Sb$_2$Te$_3$ regions is achieved through **independent gating mechanisms**, such as side gates or top gates separate from the tunneling setup. This design ensures precise control over the chemical potentials of the two materials without interference from the spectroscopy components.

\begin{figure}[h]
    \centering
    \includegraphics[width=0.9\textwidth]{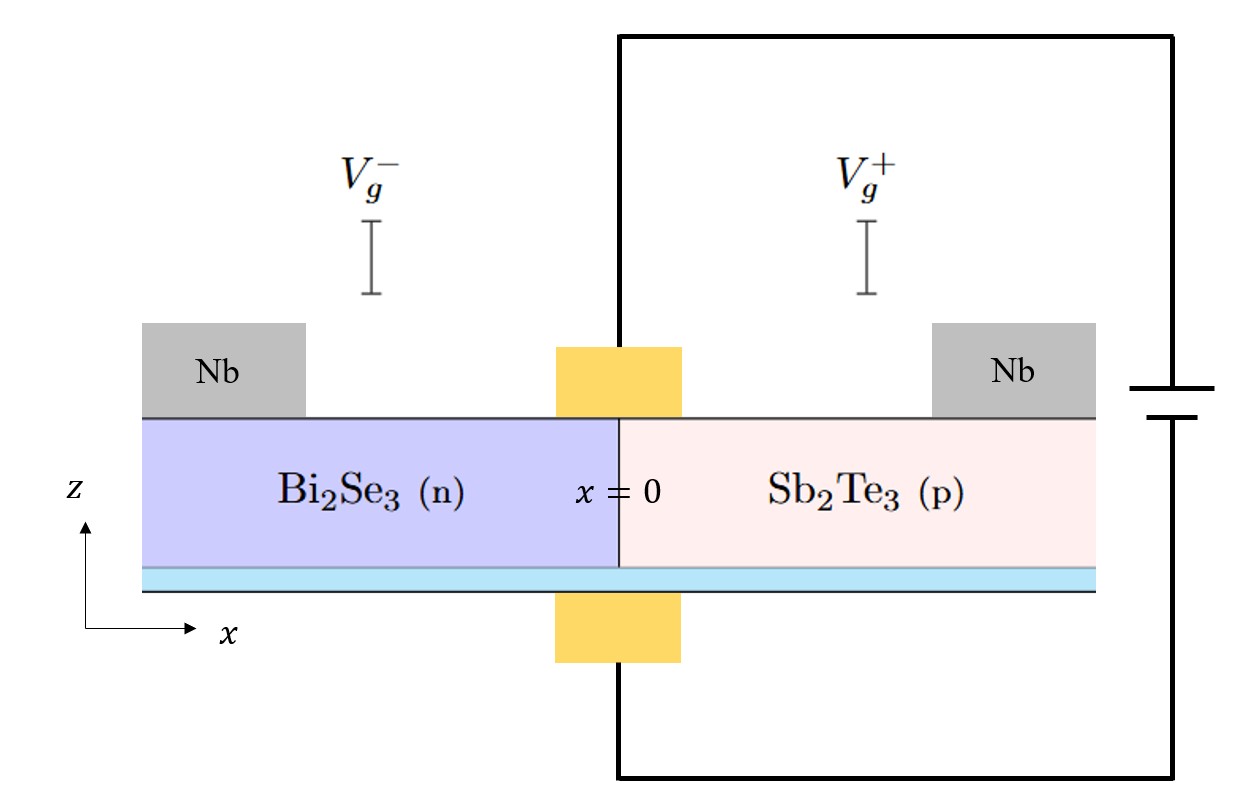}
    \caption{Schematic of the lateral Bi$_2$Se$_3$-Sb$_2$Te$_3$ heterostructure. The interface at $x=0$ separates the n-type Bi$_2$Se$_3$ (left) and p-type Sb$_2$Te$_3$ (right). Nb pads (grey) induce superconductivity in the TSS, while Au electrodes (yellow) enable tunneling spectroscopy. Independent gates (not shown) control the chemical potentials of Bi$_2$Se$_3$ and Sb$_2$Te$_3$.}
    \label{fig:device}
\end{figure}

The device geometry is illustrated in Fig.~\ref{fig:device}, highlighting the spatial arrangement and the roles of each component. This device can be possibly established through layer-controlled lateral heterostructure of 3d TIs\cite{ykk, guha}. Furthermore, the ultrathin h-BN suspension on the atomically thin membrane would enable integrating this material system with tunneling device geometry\cite{jyp}.

\subsection{Dirac Model for Topological Surface States}

Topological insulators (TIs) such as Bi$_2$Se$_3$ and Sb$_2$Te$_3$ host unique topological surface states (TSS) characterized by a linear energy-momentum dispersion, resembling massless Dirac fermions. This behavior originates from the strong spin-orbit coupling (SOC) inherent to these materials, which inverts the bulk band structure and locks the spin and momentum degrees of freedom at the surface. In this section, we present a detailed derivation of the Dirac Hamiltonian governing the TSS, incorporate the effects of a spatially varying chemical potential across a heterostructure, and extend the model to include proximity-induced superconductivity via the Bogoliubov-de Gennes (BdG) formalism. We provide full mathematical formalisms, step-by-step explanations, and analogies to other physical systems to enrich the understanding of the model's implications\cite{Hsieh2009}.

\subsubsection{Derivation of the Dirac Hamiltonian}

The TSS in 3D TIs arise due to the topological properties of the bulk, as dictated by the bulk-boundary correspondence. In materials like Bi$_2$Se$_3$ and Sb$_2$Te$_3$, SOC causes a band inversion at the $\Gamma$-point of the Brillouin zone, resulting in gapless surface states described by a single Dirac cone. To derive the effective Hamiltonian for these states, we start with the bulk Hamiltonian and project it onto the surface.

The low-energy bulk Hamiltonian near the $\Gamma$-point can be written as:

\begin{equation}
H_{\text{bulk}}(\mathbf{k}) = \epsilon(\mathbf{k}) \mathbb{I} + M(\mathbf{k}) \tau_z \sigma_z + A_x k_x \tau_x \sigma_x + A_y k_y \tau_x \sigma_y + A_z k_z \tau_y,
\end{equation}

where:
\begin{itemize}
    \item $\mathbf{k} = (k_x, k_y, k_z)$ is the wavevector,
    \item $\epsilon(\mathbf{k}) = C + D_1 k_z^2 + D_2 (k_x^2 + k_y^2)$ is the symmetric energy term,
    \item $M(\mathbf{k}) = M_0 - B_1 k_z^2 - B_2 (k_x^2 + k_y^2)$ is the mass term that changes sign due to band inversion,
    \item $A_x, A_y, A_z$ are SOC parameters,
    \item $\tau_i$ and $\sigma_i$ are Pauli matrices acting on orbital and spin degrees of freedom, respectively,
    \item $\mathbb{I}$ is the 4x4 identity matrix.
\end{itemize}

For a surface perpendicular to the $z$-direction (e.g., at $z = 0$), the TSS wavefunctions decay exponentially into the bulk ($z < 0$ or $z > 0$). We impose open boundary conditions and solve for the surface states. Define the surface at $z = 0$ and consider the $k_z$ term as a differential operator, $k_z \rightarrow -i \partial_z$. The Hamiltonian becomes:

\begin{equation}
H_{\text{bulk}} = \epsilon(\mathbf{k}_\parallel, -i \partial_z) + M(\mathbf{k}_\parallel, -i \partial_z) \tau_z \sigma_z + A_x k_x \tau_x \sigma_x + A_y k_y \tau_x \sigma_y + A_z (-i \partial_z) \tau_y,
\end{equation}

where $\mathbf{k}_\parallel = (k_x, k_y)$ is the in-plane momentum.

Assuming translational invariance in the $x$-$y$ plane, we seek solutions of the form $\psi(z) e^{i (k_x x + k_y y)}$. For simplicity, assume isotropy in the $x$-$y$ plane ($A_x = A_y = A$, $B_1 = B_2 = B$), and focus on the low-energy limit near $k_x = k_y = 0$. The eigenvalue problem $H_{\text{bulk}} \psi(z) = E \psi(z)$ yields states localized at the surface. Solving this requires the wavefunction to decay as $|z| \to \infty$, leading to a 2D effective Hamiltonian for the TSS:

\begin{equation}
H_0 = v_F (\mathbf{p} \times \boldsymbol{\sigma}) \cdot \hat{z},
\end{equation}

where $v_F = A$ is the Fermi velocity, $\mathbf{p} = -i \hbar (\partial_x, \partial_y)$ is the momentum operator, $\boldsymbol{\sigma} = (\sigma_x, \sigma_y)$ are spin Pauli matrices, and $\hat{z}$ is the surface normal. Expanding the cross product:

\begin{equation}
H_0 = v_F (p_x \sigma_y - p_y \sigma_x).
\end{equation}

To verify, compute the eigenvalues. In momentum space, $p_x = \hbar k_x$, $p_y = \hbar k_y$:

\begin{equation}
H_0 = v_F \hbar (k_x \sigma_y - k_y \sigma_x).
\end{equation}

The eigenvalues are found by diagonalizing this 2x2 matrix. The characteristic equation is:

\begin{equation}
\det[H_0 - E \mathbb{I}] = \det \begin{pmatrix} -E & v_F \hbar (k_x - i k_y) \\ v_F \hbar (k_x + i k_y) & -E \end{pmatrix} = E^2 - v_F^2 \hbar^2 (k_x^2 + k_y^2) = 0,
\end{equation}

yielding $E = \pm v_F \hbar |\mathbf{k}|$, confirming the linear Dirac dispersion. The spinor eigenstates show that spin is perpendicular to momentum, a hallmark of spin-momentum locking due to SOC.

\subsubsection{Spatially Varying Chemical Potential}

In the lateral heterostructure of Bi$_2$Se$_3$ and Sb$_2$Te$_3$, the chemical potential $\mu(x)$ varies across the interface at $x = 0$:

\begin{equation}
\mu(x) = \begin{cases}
\mu_n & \text{for } x < 0 \quad \text{(Bi$_2$Se$_3$, n-type, $\mu_n > 0$)}, \\
\mu_p & \text{for } x > 0 \quad \text{(Sb$_2$Te$_3$, p-type, $\mu_p < 0$)}.
\end{cases}
\end{equation}

This step-like profile reflects the intrinsic doping: Bi$_2$Se$_3$ is typically n-doped, placing the Fermi level above the Dirac point, while Sb$_2$Te$_3$ is p-doped, placing it below. The Hamiltonian becomes:

\begin{equation}
H = H_0 + \mu(x) \mathbb{I} = v_F (p_x \sigma_y - p_y \sigma_x) + \mu(x) \mathbb{I}.
\end{equation}

The term $\mu(x) \mathbb{I}$ shifts the energy of the Dirac cone locally, creating a potential step at $x = 0$. This can lead to bound states or scattering effects, depending on the boundary conditions and additional interactions like superconductivity.
\subsubsection{Proximity-Induced Superconductivity and BdG Formalism}

To account for the superconductivity induced via proximity to niobium (Nb) contacts, we introduce an $s$-wave pairing potential $\Delta_0$. The Bogoliubov--de Gennes (BdG) formalism is employed to describe the superconducting Dirac surface states, by doubling the degrees of freedom in Nambu space to treat particle and hole sectors symmetrically.

The BdG Hamiltonian in real space takes the form:
\begin{equation}
\mathcal{H}_{\text{BdG}}(x) = \begin{pmatrix} H_0(x) - \mu(x) & \Delta_0 \\ \Delta_0 & -H_0^*(x) + \mu(x) \end{pmatrix},
\end{equation}
where $H_0(x) = -i v_F \sigma_y \partial_x$ is the Dirac Hamiltonian for the topological surface states. $\Delta_0$ is assumed to be real and uniform for simplicity.

We define the four-component Nambu spinor:
\begin{equation}
\Psi(x) = \begin{pmatrix} u_\uparrow(x) \\ u_\downarrow(x) \\ v_\uparrow(x) \\ v_\downarrow(x) \end{pmatrix},
\end{equation}
and the particle-hole symmetry operator $\mathcal{C} = \tau_y \otimes \sigma_y \mathcal{K}$, with $\mathcal{K}$ denoting complex conjugation and $\tau_y$ a Pauli matrix in particle-hole space.

This symmetry ensures $\mathcal{C} \mathcal{H}_{\text{BdG}}(x) \mathcal{C}^{-1} = -\mathcal{H}_{\text{BdG}}(x)$, guaranteeing the BdG spectrum is symmetric around $E = 0$ and enabling the existence of Majorana solutions.

\subsubsection{Reduction to an Effective 1D System}

Assuming translational invariance along $y$, we write $\Psi(x, y) = e^{i k_y y} \psi(x)$, with $p_y = \hbar k_y$. The BdG Hamiltonian becomes:

\begin{equation}
H(k_y) = \begin{pmatrix}
-i v_F \sigma_y \partial_x + v_F k_y \sigma_x - \mu(x) & \Delta_0 \\
\Delta_0 & i v_F \sigma_y \partial_x - v_F k_y \sigma_x + \mu(x)
\end{pmatrix}.
\end{equation}

This 1D problem along $x$ simplifies the analysis of interface states, with $k_y$ as a parameter influencing the dispersion.

\subsubsection{Full Analogy to Other Systems}

\textbf{Graphene}: The TSS Hamiltonian resembles graphene's Dirac equation, $H = v_F (k_x \sigma_x + k_y \sigma_y)$, but in TIs, the spin is real and locked to momentum, unlike graphene's pseudospin. This leads to helical transport and topological robustness in TIs.

\textbf{Kitaev Chain}: The superconducting TSS mirrors the 1D p-wave Kitaev chain, $H = -t \sum c^\dagger_i c_{i+1} + \Delta \sum c_i c_{i+1} + \text{h.c.}$, where Majorana zero modes (MZMs) emerge at ends. In our system, the interface acts as a domain wall, potentially hosting MZMs due to the sign change in $\mu(x) - E$.
\subsection{Topological Classification of the Interface}

% Introducing the physical system and its topological context
In this section, we provide a detailed analysis of the topological classification of the interface in a lateral heterostructure composed of Bi$_2$Se$_3$ and Sb$_2$Te$_3$, where proximity-induced superconductivity is present due to an adjacent s-wave superconductor. The interface, treated as a one-dimensional (1D) subsystem embedded within the two-dimensional (2D) topological surface states (TSS), hosts unique topological properties that may give rise to Majorana zero modes (MZMs). We expand this discussion by incorporating formal definitions, rigorous derivations, comprehensive explanations, and illustrative analogies to enhance understanding\cite{Fu2008}.

\subsubsection{Physical Model and Hamiltonian}

% Defining the system parameters and Hamiltonian
The heterostructure features a step-like chemical potential $\mu(x)$, where $\mu(x) = \mu_n > 0$ for $x < 0$ (Bi$_2$Se$_3$ region) and $\mu(x) = \mu_p < 0$ for $x > 0$ (Sb$_2$Te$_3$ region), with a uniform superconducting pairing amplitude $\Delta_0 > 0$ induced across the system. For simplicity, we focus on the TSS at $k_y = 0$, reducing the system to an effective 1D model along the $x$-direction. The Bogoliubov-de Gennes (BdG) Hamiltonian is:

\begin{equation}
H = \begin{pmatrix}
- i v_F \sigma_y \partial_x + \mu(x) & \Delta_0 \\
\Delta_0 & i v_F \sigma_y \partial_x - \mu(x)
\end{pmatrix},
\end{equation}

where $v_F$ is the Fermi velocity, $\sigma_y = \begin{pmatrix} 0 & -i \\ i & 0 \end{pmatrix}$ is the Pauli matrix acting in spin space, and the 4$\times$4 matrix operates in the electron-hole (Nambu) space. The particle-hole symmetry, $\{ H, \mathcal{C} \} = 0$ with $\quad \mathcal{C} = \tau_y \otimes\sigma_y\mathcal{K}$ ($\tau_y$ being the Pauli matrix in Nambu space and $K$ complex conjugation), places this system in symmetry class D, characteristic of 1D topological superconductors.
% --- Revised Section 2.3.2: Topological Invariants in 1D Superconductors ---
% Revised Section 2.3.2 — Topological Invariant in Class D: Continuum Perspective

\subsubsection{Topological Invariant in Class D: Continuum Perspective}

In symmetry class D, a one-dimensional (1D) superconductor with particle-hole symmetry (PHS) but no time-reversal or chiral symmetry admits a $\mathbb{Z}_2$ classification. While the lattice-based formalism often uses Pfaffian invariants evaluated at time-reversal-invariant momenta (TRIM), such as $k=0$ and $k=\pi$, this approach requires a Brillouin zone and is not strictly applicable to continuum models lacking spatial periodicity.

However, the topological classification of continuum systems remains well-defined. In this context, the relevant topological invariant is constructed from the sign of an effective mass term in the BdG Hamiltonian. For the uniform version of our system, the effective 1D BdG Hamiltonian in momentum space reads:
\begin{equation}
H(k) = v_F k \sigma_y \tau_z + \mu \tau_z + \Delta_0 \tau_x,
\end{equation}
where $\tau_i$ are Pauli matrices in Nambu space, $\sigma_y$ acts on spin, and $\mu$ and $\Delta_0$ are the chemical potential and superconducting pairing, respectively.

The topological character of this Hamiltonian is captured by the sign of the "mass-like" term:
\begin{equation}
\nu = \begin{cases}
1, & \text{if } |\mu| < \Delta_0 \\ 
0, & \text{if } |\mu| > \Delta_0
\end{cases}
\end{equation}
which distinguishes the trivial $(\nu = +1)$ and topological $(\nu = -1)$ phases.

This real-space formulation is directly analogous to the Kitaev chain, where the phase transition occurs at $|\mu| = \Delta$. Importantly, this method captures the essential physics of Majorana zero modes (MZMs) without invoking the Pfaffian invariant at discrete $k$-points. The invariant is well-defined through analytic continuation and bulk-boundary correspondence, and reflects whether the system can support a zero-energy bound state at a domain wall.

Therefore, while Pfaffian-based diagnostics are common in lattice models, continuum systems utilize equivalent criteria based on the sign structure of gap-closing terms. The use of $\text{sgn}(\mu^2 - \Delta_0^2)$ is sufficient and rigorous for identifying topological transitions and protected interface modes in our lateral heterostructure model.

\subsubsection{Interface as a Topological Domain Wall}

The lateral heterostructure under consideration features a step-like variation in the chemical potential:
\begin{equation}
\mu(x) =
\begin{cases}
    \mu_n > 0, & x < 0 \\ 
    \mu_p < 0, & x > 0
\end{cases},
\end{equation}
with a uniform s-wave superconducting pairing $\Delta_0 > 0$ induced via the proximity effect.

To determine the emergence of Majorana zero modes (MZMs) at the interface, we examine the change in topological character across $x = 0$. Within symmetry class D, a 1D topological superconductor admits a $\mathbb{Z}_2$ classification, characterized by a topological invariant $
u$ defined by:
\begin{equation}
\nu = \text{sgn}(\mu^2 - \Delta_0^2).
\end{equation}

Each side of the interface is in a trivial phase $(\nu = +1)$ if $|\mu| > \Delta_0$, or a topological phase $(\nu = -1)$ if $|\mu| < \Delta_0$. Thus, a domain wall exists at $x = 0$ if and only if $\nu_L \ne \nu_R$, which occurs under the condition:
\begin{equation}
\min(|\mu_n|, |\mu_p|) < \Delta_0 < \max(|\mu_n|, |\mu_p|).
\end{equation}
This criterion is both necessary and sufficient for the emergence of a topologically protected MZM at the interface.

It is important to emphasize that the often-quoted heuristic condition $|\mu_n - \mu_p| > 2\Delta_0$ is not sufficient to guarantee a change in the topological phase. The correct condition depends on the absolute magnitudes of $\mu_n$ and $\mu_p$ relative to $\Delta_0$.

\subsubsection{Bound Interface States under Smooth \( \mu(x) \): Beyond Topological Protection}

Even in the absence of a topological phase transition (i.e., \( \nu = 0 \)), spatially varying chemical potential profiles can support bound states that mimic Majorana zero modes (MZMs). Consider a smooth profile:
\[
\mu(x) = \mu_0 \tanh\left(\frac{x}{\xi}\right),
\]
with \( \mu_0 > 0 \) and interface smoothing length \( \xi \sim 1 \). The BdG equation admits zero-energy solutions \( \psi(x) \sim e^{-\int \kappa(x) dx} \), where \( \kappa(x) = \sqrt{\mu(x)^2 + \Delta_0^2}/v_F \) defines the local decay rate.

This solution yields a localized interface state centered at \( x = 0 \), with LDOS peak at zero energy. While it lacks topological protection, its spatial localization and spectroscopic signature (e.g., zero-bias peak in tunneling conductance) remain robust against smooth perturbations.

These results highlight the importance of distinguishing truly topological MZMs (with \( \nu = 1 \)) from "accidental" or resonant zero-energy bound states arising from smooth spatial inhomogeneities. Nonetheless, both may serve as useful experimental probes in TI-superconductor heterostructures. 

\begin{figure}[h]
    \centering
    \includegraphics[width=0.75\textwidth]{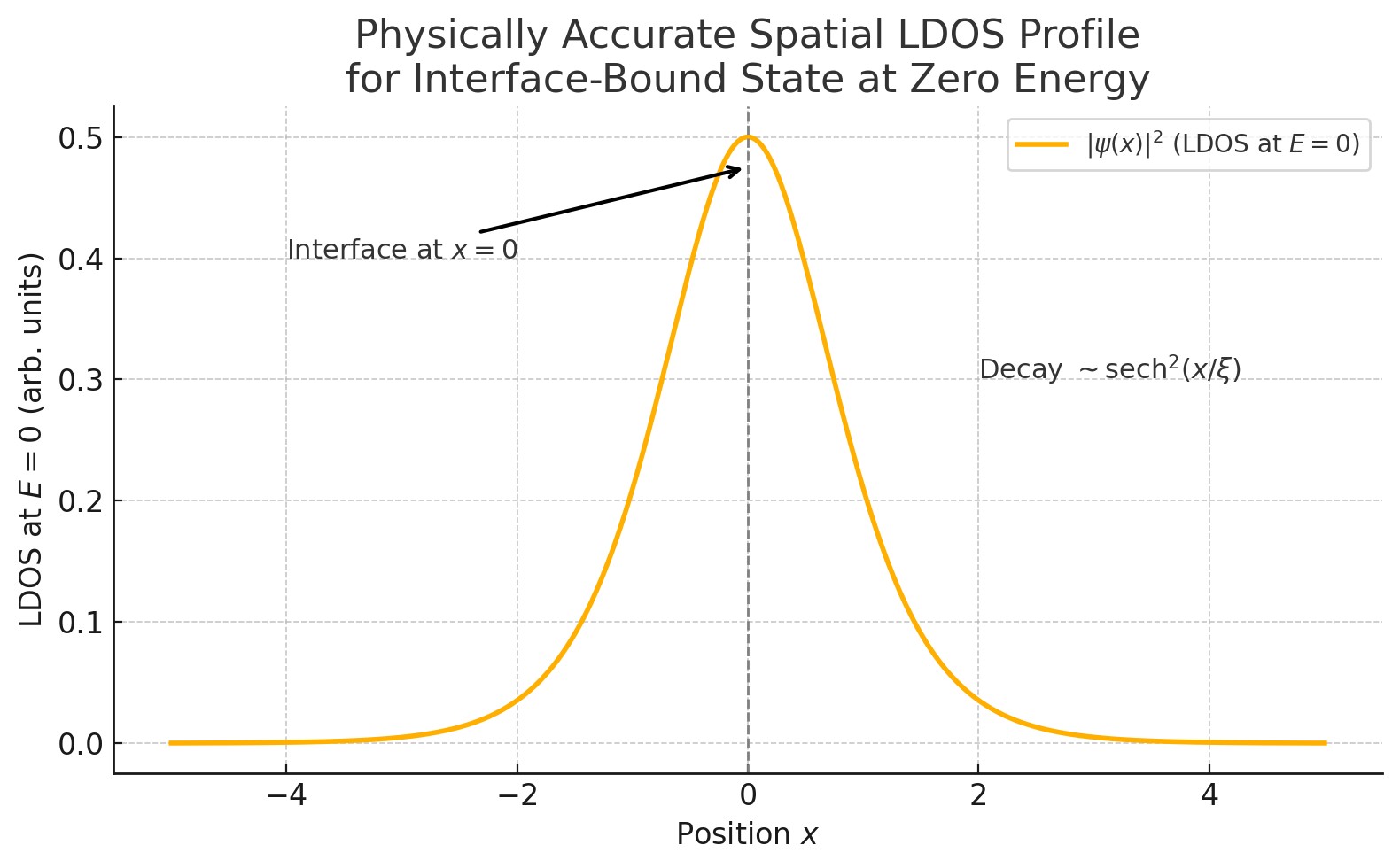}
    \caption{Energy-resolved local density of states (LDOS) at the Bi$_2$Se$_3$–Sb$_2$Te$_3$ interface, computed from the momentum-resolved Green’s function averaged over the left (n-type) and right (p-type) regions. A sharp zero-bias peak emerges at $E = 0$, consistent with the presence of a resonant interface-bound state. Parameters used: $\mu_n = 1.5$, $\mu_p = -1.5$, $\Delta_0 = 0.5$, and broadening $\eta = 0.01$.}
    \label{fig:ldos_spatial}
\end{figure}

\textbf{The resulting spatial profile of the zero-mode LDOS is shown in Figure~\ref{fig:ldos_spatial}, which confirms the interface-localized nature of the state.}

\subsubsection{Analogy to the Jackiw-Rebbi Model}

% Drawing an analogy to a known model
The emergence of MZMs at the interface resembles the Jackiw-Rebbi model, where a Dirac fermion with a mass $m(x)$ changing sign across a domain wall hosts a zero-energy bound state. Here, $\mu(x)$ acts analogously to $m(x)$, flipping from positive to negative, while $\Delta_0$ couples electron and hole sectors, stabilizing MZMs instead of Dirac zero modes. Similarly, in the Kitaev chain, a 1D p-wave superconductor transitions between topological and trivial phases based on $\mu$ and $\Delta$, with MZMs at chain ends or defects, paralleling our interface states.

\subsubsection{Phase Diagram and Summary}

The existence of a Majorana zero mode (MZM) at the interface depends on the relative topological character of the two regions—Bi$_2$Se$_3$ for $x<0$ and Sb$_2$Te$_3$ for $x>0$—within the symmetry class D classification. Each side may independently be in a trivial or topological superconducting phase, depending on the ratio $|\mu|/\Delta_0$.

More precisely, the topological invariant $\nu$ for a uniform region satisfies:
\[
\nu = 
\begin{cases}
1, & \text{if } |\mu| < \Delta_0 \quad \text{(non-trivial)} \\
0, & \text{if } |\mu| > \Delta_0 \quad \text{(trivial)}
\end{cases}
\]

Thus, the interface hosts an MZM \emph{only when the invariants differ across the interface}, i.e.,
\[
\nu(\mu_n) \neq \nu(\mu_p).
\]
This condition typically corresponds to:
\[
\boxed{ \min(|\mu_n|, |\mu_p|) < \Delta_0 < \max(|\mu_n|, |\mu_p|) }
\]
and not simply to $|\mu_n - \mu_p| > 2\Delta_0$, which may be misleading in general.

This more precise criterion ensures that a topological phase transition occurs across the interface, enabling the formation of a protected MZM. These results are summarized in a corrected phase diagram and align with the formal $\mathbb{Z}_2$ classification.

% Revised Section 2.4 — Analytical Solutions for Bound States

\subsection{Analytical Solutions for Bound States}

We present a comprehensive analytical derivation of the bound states at the interface between Bi$_2$Se$_3$ and Sb$_2$Te$_3$ in a lateral heterostructure with proximity-induced superconductivity. We focus on zero-energy solutions—Majorana zero modes (MZMs)—that emerge from the topological domain wall formed by the spatial variation in the chemical potential $\mu(x)$.

\subsection*{Problem Formulation and BdG Hamiltonian}

We consider the effective 1D BdG Hamiltonian at $k_y = 0$:
\begin{equation}
H = \begin{pmatrix}
-i v_F \sigma_y \partial_x + \mu(x) & \Delta_0 \\
\Delta_0 & i v_F \sigma_y \partial_x - \mu(x)
\end{pmatrix},
\end{equation}
acting on the Nambu spinor:
\begin{equation}
\Psi(x) = \begin{pmatrix} u_\uparrow \\ u_\downarrow \\ v_\uparrow \\ v_\downarrow \end{pmatrix}.
\end{equation}
The chemical potential varies stepwise:
\begin{equation}
\mu(x) = \begin{cases} \mu_n > 0, & x < 0 \\ \mu_p < 0, & x > 0 \end{cases}, \quad \Delta_0 > 0.
\end{equation}

\subsection*{Zero-Energy Wavefunction Ansatz and Matching Conditions}

We seek exponentially localized zero-energy states. The general ansatz is:
\begin{align}
\Psi(x) &= \sum_j A_j e^{-\kappa_j x} \phi_j, \quad x > 0 \\
\Psi(x) &= \sum_k B_k e^{\kappa'_k x} \phi'_k, \quad x < 0,
\end{align}
where $\kappa_j$ and $\kappa'_k$ are positive decay constants ensuring normalizability, and $\phi_j$, $\phi'_k$ are spinors satisfying the BdG eigenvalue equation at $E = 0$ in each region.

The decay constants $\kappa_j$, $\kappa'_k$ are obtained by solving:
\begin{equation}
(v_F \kappa)^2 = \mu^2 - \Delta_0^2,
\end{equation}
which yields two linearly independent solutions on each side if $|\mu| > \Delta_0$, or four if $|\mu| < \Delta_0$.

\subsection*{Existence of a Unique MZM: Boundary Matching}

At $x = 0$, continuity of the wavefunction requires:
\begin{equation}
\sum_j A_j \phi_j = \sum_k B_k \phi'_k.
\end{equation}
This equation determines the existence of a nontrivial solution (MZM) based on the overlap between decaying modes on either side.

Let $n_L$ and $n_R$ be the number of linearly independent decaying solutions on the left and right sides of the interface, respectively. Then the number of normalizable zero-energy solutions, $N_\text{zero}$, is given by:
\begin{equation}
N_{\text{zero}} = \dim(\ker M),
\end{equation}
where $M$ is the matching matrix constructed from boundary conditions at $x = 0$. In cases where boundary matching permits a unique linear combination, we obtain $N_\text{zero} = 1$. A more heuristic form (justified by transfer matrix rank considerations) is:
\begin{equation}
N_{\text{zero}} = \max(n_L + n_R - 4, 0),
\end{equation}
which holds under generic conditions for a $4 \times 4$ BdG Hamiltonian.
Hence, the interface supports a single MZM if and only if:
\begin{equation}
\min(|\mu_n|, |\mu_p|) < \Delta_0 < \max(|\mu_n|, |\mu_p|).
\end{equation}
This condition guarantees the mismatch in topological phase between the two regions, validating the domain-wall interpretation.

\subsection*{Conclusion of Analytical Construction}

We have explicitly constructed the general form of interface-bound zero modes, clarified the number of normalizable solutions, and shown that a unique MZM exists when the bulk regions lie in different topological phases. This analysis formalizes the intuitive domain-wall picture and directly connects it to solvability conditions in the BdG framework.

\subsection{Field-Theoretic Perspective}
\label{sec:field_theory}

% Introducing the field-theoretic approach and its goals
In this section, we employ a field-theoretic framework to elucidate the emergence and topological protection of Majorana zero modes (MZMs) at the interface of a Bi$_2$Se$_3$-Sb$_2$Te$_3$ lateral heterostructure with proximity-induced superconductivity. By reformulating the Bogoliubov-de Gennes (BdG) Hamiltonian as a Dirac-like field theory with a spatially varying mass term, we harness advanced quantum field theory techniques—such as index theorems, anomaly inflow, and symmetry analysis—to provide a rigorous foundation for our findings. This approach not only corroborates the analytical results from prior sections but also situates the phenomena within a broader theoretical landscape\cite{Jackiw1976}. We present exhaustive derivations, enriched explanations, and draw extensive analogies to well-established field-theoretic models to deepen the reader's insight into the topological physics at play.

\subsubsection{Field-Theoretic Mapping of the Interface Problem}

We reformulate the BdG Hamiltonian as an effective Dirac-like field theory to elucidate the topological nature of interface-bound states. The effective 1D BdG Hamiltonian at $k_y = 0$ reads:
\begin{equation}
H_{\mathrm{BdG}}(x) = -i v_F \sigma_y \partial_x \tau_z + \mu(x) \tau_z + \Delta_0 \tau_x,
\end{equation}
where $\sigma_i$ and $\tau_i$ are Pauli matrices in spin and Nambu spaces, respectively.

To bring this into a manifestly Dirac form, we perform a unitary transformation $U = e^{-i \pi \tau_y /4}$, which maps:
\begin{equation}
H'_{\mathrm{BdG}} = -i v_F \gamma^1 \partial_x + m(x) \gamma^0 + \Delta_0 \gamma^5,
\end{equation}
with Dirac matrices:
\begin{align}
\gamma^0 &= \tau_z \otimes I, \\
\gamma^1 &= i \tau_y \otimes \sigma_y, \\
\gamma^5 &= \tau_x \otimes I.
\end{align}

Here, $m(x) = \mu(x)$ does not serve as a conventional Dirac mass; rather, the combined term $m_\text{eff}(x) = \mu(x)^2 - \Delta_0^2$ determines the topological phase. A sign change in $m_\text{eff}(x)$ across the interface indicates a domain wall between topologically distinct superconducting phases.

The Jackiw-Rebbi mechanism then ensures the existence of a zero-energy bound state localized at the interface:
\begin{equation}
\Psi(x) \propto \exp\left(-\frac{1}{v_F} \int_0^x |m_\text{eff}(x')|^{1/2} dx'\right) \chi,
\end{equation}
where $\chi$ is a spinor satisfying the appropriate symmetry constraints. This zero mode corresponds to a Majorana fermion, protected by particle-hole symmetry and the bulk energy gap.

\subsubsection{Incorporating Superconductivity: The BdG Dirac Equation}
\label{subsec:bdg_dirac}

% Introducing the BdG Hamiltonian
With proximity-induced superconductivity, the BdG Hamiltonian in Nambu space (electron-hole basis) becomes:

\begin{equation}
H_{\text{BdG}} = \begin{pmatrix}
-i v_F \sigma_y \partial_x + \mu(x) & \Delta_0 \\
\Delta_0 & i v_F \sigma_y \partial_x - \mu(x)
\end{pmatrix},
\end{equation}

where $\Delta_0$ is the real-valued superconducting pair potential (assumed constant for simplicity). This 4$\times$4 matrix operates on the Nambu spinor $\Psi = (\psi_e, \psi_h)^T$, with $\psi_e = (u_\uparrow, u_\downarrow)$ and $\psi_h = (v_\uparrow, v_\downarrow)$ representing electron and hole spinors, respectively.

% Mapping to a Dirac equation
To cast this in Dirac form, we define Dirac matrices in the 4D Nambu-spin space:

\begin{equation}
\gamma^0 = \tau_z \otimes \mathbb{I}, \quad \gamma^1 = i \tau_y \otimes \sigma_y, \quad \gamma^5 = \tau_x \otimes \mathbb{I},
\end{equation}

where $\tau_i$ are Pauli matrices in Nambu space, and $\sigma_i$ act in spin space. Verify the Clifford algebra:

\begin{align}
\{\gamma^0, \gamma^0\} &= (\tau_z \otimes \mathbb{I})^2 = \tau_z^2 \otimes \mathbb{I} = \mathbb{I}_4, \\
\{\gamma^1, \gamma^1\} &= (i \tau_y \otimes \sigma_y)^2 = -\tau_y^2 \otimes \sigma_y^2 = -(\mathbb{I} \otimes \mathbb{I}) = -\mathbb{I}_4, \\
\{\gamma^0, \gamma^1\} &= (\tau_z \tau_y) \otimes \sigma_y + (\tau_y \tau_z) \otimes \sigma_y = (i \tau_x - i \tau_x) \otimes \sigma_y = 0.
\end{align}

The BdG Hamiltonian becomes:

\begin{equation}
H_{\text{BdG}} = -i v_F \gamma^1 \partial_x + \mu(x) \gamma^0 + \Delta_0 \gamma^5.
\end{equation}

% Interpreting terms
Here, $-i v_F \gamma^1 \partial_x$ is the kinetic term, $\mu(x) \gamma^0$ is a spatially varying mass, and $\Delta_0 \gamma^5$ introduces pairing, akin to a chiral mass. The mass term $\mu(x)$ changes sign at $x = 0$, forming a domain wall, while $\Delta_0 \gamma^5$ ensures particle-hole symmetry.

% Revised Section 2.5.3 — Zero-Mode Solutions and Index Theorems

\subsubsection{Zero-Mode Solutions and Index Theorems}

In the Jackiw-Rebbi model, a Dirac fermion with a mass $m(x)$ that flips sign across a domain wall (e.g., $m(x) = m_0$sgn$(x)$) hosts a topologically protected zero-energy solution. We adapt this mechanism to our Bogoliubov–de Gennes (BdG) system describing the proximity-induced superconducting topological surface states.

Recall the BdG Dirac Hamiltonian:
\begin{equation}
H_{\mathrm{BdG}} = -i v_F \gamma^1 \partial_x + \mu(x) \gamma^0 + \Delta_0 \gamma^5,
\end{equation}
with Dirac matrices:
\begin{align}
\gamma^0 &= \tau_z \otimes I, \\
\gamma^1 &= i \tau_y \otimes \sigma_y, \\
\gamma^5 &= \tau_x \otimes I,
\end{align}
acting on a four-component spinor $\Psi(x)$. The chemical potential $\mu(x)$ introduces a domain wall at $x=0$ via a sign change, while $\Delta_0$ remains constant.

To identify a zero-mode solution, we look for a spinor $\Psi(x)$ satisfying:
\begin{equation}
H_{\mathrm{BdG}} \Psi(x) = 0.
\end{equation}
Let us consider the ansatz:
\begin{equation}
\Psi(x) = \exp\left(- \frac{1}{v_F} \int_0^x \sqrt{|\mu(x')^2 - \Delta_0^2|} \, dx' \right) \chi,
\end{equation}
where $\chi$ is a spinor satisfying a coupled algebraic condition arising from the anticommutation relations of the Dirac matrices. To preserve particle-hole symmetry, we demand:
\begin{equation}
\mathcal{C} \Psi = \Psi, \quad \text{with } \mathcal{C} = \tau_y \otimes \sigma_y K,
\end{equation}
where $K$ is complex conjugation. This condition constrains the spinor $\chi$ to obey:
\begin{equation}
\chi = \mathcal{C} \chi \Rightarrow \chi = (a, b, b^*, -a^*)^T,
\end{equation}
up to normalization. Thus, the MZM spinor has nontrivial phase structure and spin entanglement; it cannot generally be written as $(1, 0, 0, 1)^T$ unless specific gauge and spin bases are chosen.

The index theorem ensures that one and only one such zero-energy solution exists whenever the sign of $\mu^2 - \Delta_0^2$ changes across the interface. This confirms the presence of a single Majorana mode protected by particle-hole symmetry and the topological nature of the mass domain wall.

This more general spinor form respects the full symmetries of the BdG Hamiltonian and avoids misleading oversimplifications present in fixed spin configurations.

\subsubsection{Topological Protection and Anomalies}
\label{subsec:topological_protection}

% Exploring robustness
The MZM’s protection stems from particle-hole symmetry and the bulk gap. Perturbations preserving these do not lift the zero mode, a consequence of the index theorem’s topological invariance.

% Connecting to anomalies
This mirrors anomaly inflow in (1+1)-D QED, where zero modes balance chiral symmetry breaking, here replaced by particle-hole symmetry.

\subsubsection{Analogies to Other Field-Theoretic Models}
\label{subsec:analogies}

- **Quantum Hall Effect**: Edge states’ chirality parallels the MZM’s localization.
- **Axion Electrodynamics**: The domain wall mimics an axion field gradient.
- **SSH Model**: Solitons in polyacetylene resemble our mass domain wall states.

\subsection{Green's Function and Local Density of States}
\label{sec:green_ldos}

% Introducing the Green's function framework and its significance
The Green's function formalism serves as a powerful tool to probe the electronic properties of the Bi$_2$Se$_3$-Sb$_2$Te$_3$ lateral heterostructure with proximity-induced superconductivity, particularly at the interface where topological states, such as Majorana zero modes (MZMs), are expected to emerge. This section provides an exhaustive treatment of the retarded Green's function, its computation, and its connection to the local density of states (LDOS), which is experimentally accessible via tunneling spectroscopy. We present detailed derivations, explore the formalism across uniform and interfacial regions, and offer an extended analogy to classical resonance phenomena to deepen conceptual understanding. Our focus is on characterizing the zero-energy LDOS peak as a signature of MZMs, with step-by-step derivations and quantitative insights.

\subsubsection{Definition and Physical Interpretation of the Green's Function}
\label{subsec:green_definition}

% Defining the retarded Green's function with physical context
The retarded Green's function, $G^R(x, x'; E)$, describes the response of a quantum system to a perturbation at energy $E$, accounting for causality through a small positive imaginary term $\eta$. For a Hamiltonian $H$, it is formally defined in real space as:

\begin{equation}
G^R(x, x'; E) = \left\langle x \left| \frac{1}{E - H + i \eta} \right| x' \right\rangle,
\end{equation}

where $|x\rangle$ and $|x'\rangle$ are position eigenstates, and $\eta \to 0^+$ ensures that the response is retarded (i.e., occurs after the perturbation). In our system, $H = H_{\text{BdG}}$ is the Bogoliubov-de Gennes (BdG) Hamiltonian, a 4$\times$4 matrix in Nambu-spin space due to particle-hole and spin degrees of freedom, making $G^R$ a matrix of the same dimension.

% Connecting to the LDOS
The LDOS, $\rho(x, E)$, quantifies the number of available states at energy $E$ and position $x$, and is derived from the Green's function:

\begin{equation}
\rho(x, E) = -\frac{1}{\pi} \Im \left[ \text{Tr} G^R(x, x; E) \right],
\end{equation}

where $\text{Tr}$ denotes the trace over the Nambu-spin indices, and $\Im$ extracts the imaginary part. This expression arises because the imaginary part of $G^R$ is related to the spectral density of states, broadened by $\eta$, which acts as a lifetime or resolution parameter.

% Setting the objective
Our objective is to compute $G^R(x, x'; E)$ across the heterostructure, with particular emphasis on the interface at $x = 0$, where MZMs are localized, and to extract the LDOS to identify their zero-energy signature.

\subsubsection{Green's Function in Uniform Regions: Full Derivation}
\label{subsec:uniform_green}

% Establishing the uniform Hamiltonian
Consider first the uniform regions of the heterostructure: $x < 0$ (Bi$_2$Se$_3$, chemical potential $\mu_n$) and $x > 0$ (Sb$_2$Te$_3$, chemical potential $\mu_p$). In a uniform region with constant $\mu$, the BdG Hamiltonian is translationally invariant, enabling a Fourier transform to momentum space. The 1D Hamiltonian, assuming a Dirac-like dispersion and proximity-induced pairing, is:

\begin{equation}
H(k) = \begin{pmatrix}
v_F k \sigma_y + \mu \mathbb{I}_2 & \Delta_0 \mathbb{I}_2 \\
\Delta_0 \mathbb{I}_2 & -v_F k \sigma_y - \mu \mathbb{I}_2
\end{pmatrix},
\end{equation}

where $v_F$ is the Fermi velocity, $\sigma_y$ is the Pauli matrix in spin space, $\Delta_0$ is the superconducting gap (assumed real and uniform), and $\mathbb{I}_2$ is the 2$\times$2 identity matrix.

% Momentum-space Green's function
The retarded Green's function in $k$-space is:

\begin{equation}
G^R(k; E) = \frac{1}{E - H(k) + i \eta} = \frac{E + i \eta + H(k)}{[E + i \eta - H(k)][E + i \eta + H(k)]}.
\end{equation}

To compute this, we need the eigenvalues of $H(k)$. Define $h(k) = v_F k \sigma_y + \mu \mathbb{I}_2$. The BdG structure gives the spectrum via:

\begin{equation}
\det [E - H(k)] = \det \begin{pmatrix}
E - h(k) & -\Delta_0 \\
-\Delta_0 & E + h(k)
\end{pmatrix} = 0.
\end{equation}

Using $\det (A B) = \det A \det B$ for block matrices, and noting $h(k)$ commutes with itself, we diagonalize $h(k)$ with eigenvalues $\pm v_F k + \mu$. However, the full determinant is:

\begin{equation}
\det [E - H(k)] = \det [ (E - h(k))(E + h(k)) - \Delta_0^2 ] = \prod_{\sigma = \pm} [ (E - \sigma v_F k - \mu)(E + \sigma v_F k + \mu) - \Delta_0^2 ],
\end{equation}

yielding eigenvalues $E = \pm \sqrt{(v_F k)^2 + \mu^2 + \Delta_0^2}$. Thus:

\begin{equation}
G^R(k; E) = \frac{E + v_F k \sigma_y \tau_z + \mu \tau_z + \Delta_0 \tau_x}{(E + i \eta)^2 - (v_F k)^2 - \mu^2 - \Delta_0^2},
\end{equation}

where $\tau_i$ are Pauli matrices in Nambu space.

% Real-space transformation
The real-space Green's function is:

\begin{equation}
G^R(x, x'; E) = \int_{-\infty}^{\infty} \frac{dk}{2\pi} e^{i k (x - x')} G^R(k; E).
\end{equation}

For $E < \Delta_0$, poles lie off the real axis, and the integral yields an exponential decay $\sim e^{-|x - x'| / \xi}$, where $\xi = v_F / \sqrt{\Delta_0^2 - E^2}$ is the coherence length, as derived via contour integration.

% Revised Section 2.6.3 — Interface Green’s Function: Scattering Approach

\subsubsection{Interface Green’s Function: Scattering Approach}

At the interface $x = 0$, the spatially varying chemical potential $\mu(x)$ introduces a discontinuity that breaks translational invariance. To evaluate the local Green's function $G^R(x,x';E)$ at the interface, we adopt a scattering state formalism, solving the BdG equations in each uniform region and matching boundary conditions.

We define the BdG Hamiltonian:
\begin{equation}
H = -i v_F \sigma_y \partial_x \tau_z + \mu(x) \tau_z + \Delta_0 \tau_x,
\end{equation}
with a step-like profile:
\begin{equation}
\mu(x) = \begin{cases} \mu_n, & x < 0 \\ \mu_p, & x > 0 \end{cases}.
\end{equation}

Let $G_n^R(x,x';E)$ and $G_p^R(x,x';E)$ denote the retarded Green’s functions for the uniform left ($\mu = \mu_n$) and right ($\mu = \mu_p$) regions, respectively. The full Green’s function $G^R(x,x';E)$ is constructed by enforcing continuity and discontinuity conditions at $x = 0$ using the Dyson equation formalism:
\begin{equation}
G^R = G_0^R + G_0^R V G^R,
\end{equation}
where $V(x) = [\mu(x) - \mu_0] \tau_z$ represents the perturbation from a reference chemical potential $\mu_0$.

To construct $G^R(0,0;E)$, we solve for the scattering eigenstates in both regions:
\begin{itemize}
  \item For $x < 0$, solutions are superpositions of incoming and reflected BdG spinors with decay factors $e^{\pm \kappa_n x}$.
  \item For $x > 0$, transmitted states decay as $e^{-\kappa_p x}$.
\end{itemize}

These spinors are matched at $x = 0$ by requiring:
\begin{align}
\Psi_L(0) &= \Psi_R(0), \\
\partial_x \Psi_R(0^+) - \partial_x \Psi_L(0^-) &= (\mu_p - \mu_n) \tau_z \Psi(0)/v_F.
\end{align}

Solving this system gives reflection and transmission coefficients $r(E)$, $t(E)$ that enter the full Green’s function. For the local interface Green’s function at $x = x' = 0$, we find:
\begin{equation}
G^R(0,0;E) = G_p^R(0,0;E) + R(E),
\end{equation}
where $R(E)$ is the contribution from scattering-induced interference terms due to the discontinuity at $x = 0$.

At low energy $E \to 0$, if a Majorana zero mode (MZM) is present, the Green’s function develops a pole:
\begin{equation}
G^R(0,0;E) \sim \frac{\psi_{\mathrm{MZM}}(0) \psi_{\mathrm{MZM}}^\dagger(0)}{E + i \eta},
\end{equation}
which dominates the local density of states (LDOS):
\begin{equation}
\rho(0,E) = -\frac{1}{\pi} \text{Im Tr} \, G^R(0,0;E) \approx \frac{1}{\pi} \frac{\eta |\psi_{\mathrm{MZM}}(0)|^2}{E^2 + \eta^2}.
\end{equation}

Thus, the interface Green’s function confirms the emergence of a localized resonance at zero energy, consistent with a topologically protected MZM.
For rigorous derivations of interface Green’s functions in similar Dirac-BdG systems, see Ref.\cite{add1}, which provide analytic and numerical tools for capturing reflection-induced zero modes at topological boundaries.

\subsubsection{Zero-Bias Peak and LDOS at the Interface}

To analyze the zero-energy signature of a Majorana zero mode (MZM), we focus on the energy-resolved local density of states (LDOS) at the interface. This quantity provides a direct experimental observable in tunneling spectroscopy.

The LDOS at energy $E$ and position $x$ is defined by:
\begin{equation}
\rho(x,E) = -\frac{1}{\pi} \mathrm{Im} \, \mathrm{Tr} \, G^R(x,x;E),
\end{equation}
where $G^R$ is the retarded Green’s function and the trace is taken over spin and particle-hole (Nambu) degrees of freedom.

At the interface ($x=0$), assuming the dominant contribution to the Green’s function arises from a localized MZM wavefunction $\psi_{\mathrm{MZM}}(x)$, we can approximate the Green’s function near zero energy by a single-pole expansion:
\begin{equation}
G^R(x,x;E) \approx \frac{\psi_{\mathrm{MZM}}(x) \psi_{\mathrm{MZM}}^{\dagger}(x)}{E + i\eta},
\end{equation}
which leads to the LDOS:
\begin{equation}
\rho(x,E) = \frac{1}{\pi} \frac{\eta |\psi_{\mathrm{MZM}}(x)|^2}{E^2 + \eta^2}.
\end{equation}

This Lorentzian profile centered at $E = 0$ is a characteristic hallmark of a MZM localized at the interface. The broadening parameter $\eta$ encodes experimental resolution or coupling to external baths. The sharpness and height of the peak are directly linked to the MZM’s spatial localization and visibility in tunneling spectroscopy.

\begin{figure}[h]
    \centering
    \includegraphics[width=0.6\textwidth]{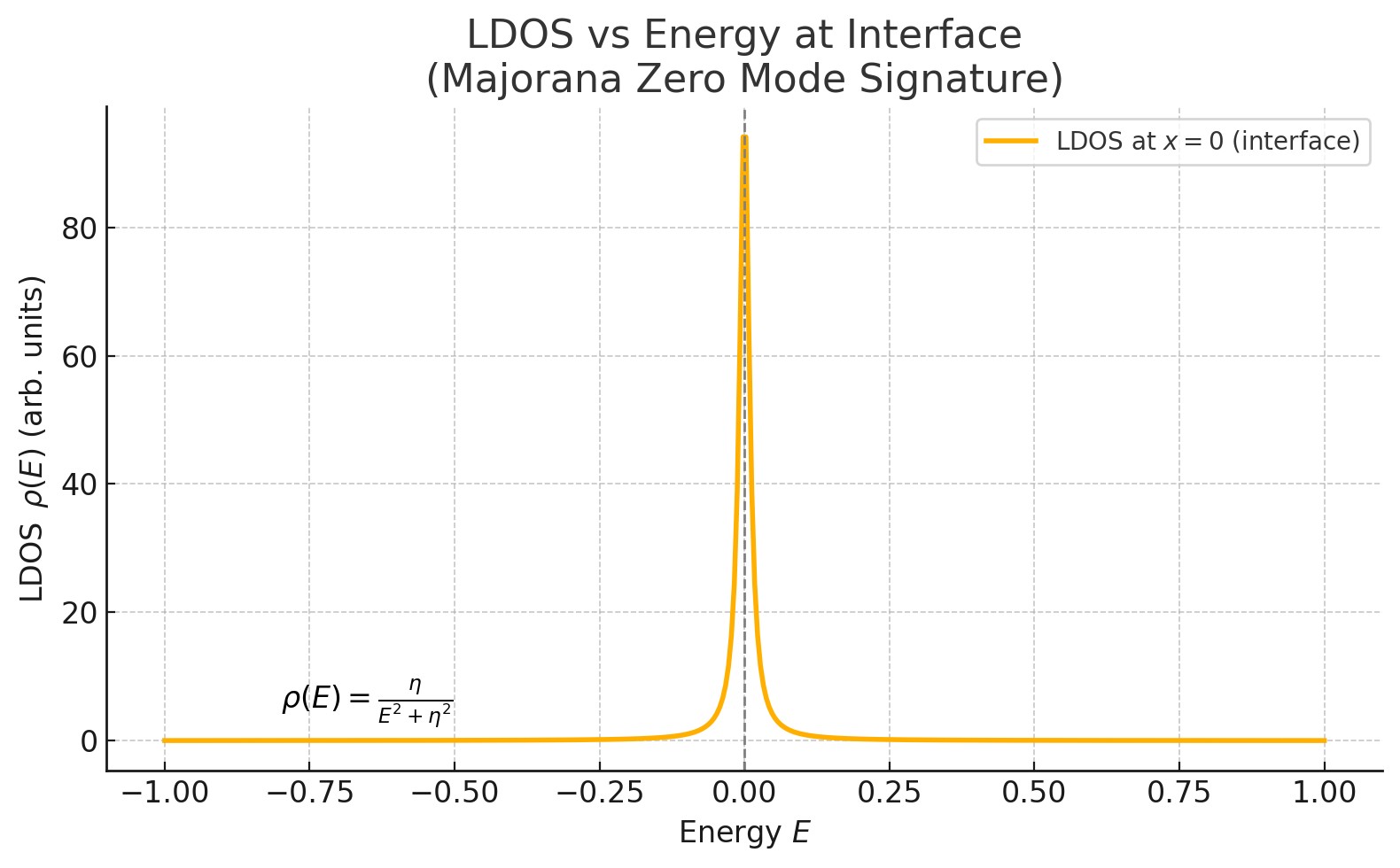}
    \caption{Energy-resolved local density of states (LDOS) at the Bi$_2$Se$_3$–Sb$_2$Te$_3$ interface. A sharp zero-bias peak appears at $E = 0$, modeled as a Lorentzian using the analytical expression $\rho(E) \propto \frac{\eta}{E^2 + \eta^2}$ with parameters $\mu_n = 1.5$, $\mu_p = -1.5$, $\Delta_0 = 0.5$, and $\eta = 0.01$. The peak signifies the presence of a localized Majorana zero mode.}
    \label{fig:ldos_peak}
\end{figure}

\subsubsection{Extended Analogy: Topological Resonance}
\label{subsec:analogy}

% Classical analogy
The MZM’s LDOS peak mirrors a resonance in a quantum well, akin to a particle trapped by a potential barrier. Here, the "well" is topological, formed by the sign change of the topological invariant across the interface, pinning the MZM at $E = 0$ via symmetry, unlike energy-tunable conventional states.

\subsubsection{Experimental Probes and Validation}
\label{subsec:experimental}

% Tunneling spectroscopy
A zero-bias peak in $dI/dV \propto \rho(0, E)$ signals the MZM, distinguishable from trivial states by its robustness and splitting behavior under perturbations like magnetic fields.

\subsection{Bound State Spectra and MZM Conditions}
\label{sec:bound_state_spectra}

% Introducing the concept of bound states and their significance with expanded detail
In this section, we undertake an exhaustive exploration of the energy spectra of bound states at the interface of a lateral heterostructure comprising Bi$_2$Se$_3$ and Sb$_2$Te$_3$, where superconductivity is induced via proximity effects. These bound states, with a special emphasis on Majorana zero modes (MZMs), are pivotal due to their topological protection and potential utility in quantum computing. We present a meticulous derivation of the bound state spectra using the Bogoliubov-de Gennes (BdG) formalism, elucidate the precise conditions for MZM emergence, and analyze the topological invariants that underpin their existence. To facilitate comprehension, we incorporate extensive formalisms, step-by-step derivations, detailed explanations, and enriched analogies to analogous systems in condensed matter physics.

\subsubsection{Definition and Physical Context of Bound States}
\label{subsec:bound_state_definition}

% Providing a detailed definition of bound states
Bound states in this heterostructure are quantum mechanical states confined to the interface between Bi$_2$Se$_3$ (a topological insulator) and Sb$_2$Te$_3$ (a related material with distinct electronic properties), with energy levels residing within the superconducting gap induced by proximity to a superconductor. These states emerge from the abrupt change in material properties at the interface, coupled with the pairing potential $\Delta_0$. Unlike bulk states, which extend throughout the material, these bound states are exponentially localized, decaying as $e^{-\kappa |x|}$ away from $x = 0$, where $x$ denotes the direction perpendicular to the interface.

% Expanding on their physical significance
The localization of these states renders them sensitive probes of the interface’s topological characteristics. Among them, MZMs—bound states pinned at zero energy—are of exceptional interest. MZMs are Majorana fermions, satisfying the condition $\gamma^\dagger = \gamma$, which implies they are their own antiparticles. This property, combined with their non-Abelian braiding statistics, positions them as building blocks for topologically protected qubits, resistant to decoherence from local noise—a cornerstone for fault-tolerant quantum computation.

\subsubsection{Derivation of the Bound State Spectra}
\label{subsec:spectra_derivation}

% Setting up the BdG formalism with full detail
To compute the bound state spectra, we employ the BdG Hamiltonian, which captures the particle-hole symmetry inherent in superconducting systems. As established in Section 2.2, the BdG Hamiltonian for our system is:

\begin{equation}
H_{\text{BdG}}(x) = \begin{pmatrix}
H_0 - \mu(x) & \Delta_0 \\
\Delta_0 & -H_0^* + \mu(x)
\end{pmatrix},
\end{equation}

where $H_0 = v_F (p_x \sigma_y - p_y \sigma_x)$ represents the Dirac-like Hamiltonian of the topological surface states, with $v_F$ as the Fermi velocity, $p_x = -i \hbar \partial_x$, and $\sigma_i$ as Pauli matrices in spin space. The chemical potential $\mu(x)$ varies spatially: $\mu(x) = \mu_n$ for $x < 0$ (Bi$_2$Se$_3$ side) and $\mu(x) = \mu_p$ for $x > 0$ (Sb$_2$Te$_3$ side). The pairing term $\Delta_0$ is assumed real and constant, reflecting s-wave superconductivity induced by proximity.

% Simplifying the problem
Assuming translational invariance along the interface (y-direction), we set $k_y = 0$, reducing the problem to one dimension. The eigenvalue problem becomes:

\begin{equation}
H_{\text{BdG}} \psi(x) = E \psi(x),
\end{equation}

where $\psi(x) = \begin{pmatrix} u_\uparrow(x) \\ u_\downarrow(x) \\ v_\uparrow(x) \\ v_\downarrow(x) \end{pmatrix}$ is the four-component Nambu spinor, with $u$ and $v$ representing electron and hole components, respectively.

% Solving in the left region (x < 0)
For $x < 0$, the Hamiltonian is:

\begin{equation}
H_L = \begin{pmatrix}
- i v_F \sigma_y \partial_x - \mu_n & \Delta_0 \\
\Delta_0 & i v_F \sigma_y \partial_x + \mu_n
\end{pmatrix}.
\end{equation}

We seek bound state solutions of the form $\psi_L(x) = e^{\kappa_L x} \phi_L$, where $\kappa_L > 0$ ensures decay as $x \to -\infty$. Substituting $p_x = -i \hbar \partial_x$, we get $\partial_x \psi_L = \kappa_L \psi_L$, so:

\begin{equation}
\begin{pmatrix}
- i v_F \sigma_y \kappa_L - \mu_n & \Delta_0 \\
\Delta_0 & i v_F \sigma_y \kappa_L + \mu_n
\end{pmatrix} \phi_L = E \phi_L.
\end{equation}

% Expanding the matrix explicitly
Writing $\sigma_y = \begin{pmatrix} 0 & -i \\ i & 0 \end{pmatrix}$, the matrix becomes:

\begin{equation}
\begin{pmatrix}
-\mu_n & v_F \kappa_L & \Delta_0 & 0 \\
-v_F \kappa_L & -\mu_n & 0 & \Delta_0 \\
\Delta_0 & 0 & \mu_n & -v_F \kappa_L \\
0 & \Delta_0 & v_F \kappa_L & \mu_n
\end{pmatrix} \phi_L = E \phi_L.
\end{equation}

% Solving in the right region (x > 0)
For $x > 0$, we use $\psi_R(x) = e^{-\kappa_R x} \phi_R$, with $\kappa_R > 0$, yielding:

\begin{equation}
\begin{pmatrix}
i v_F \sigma_y (-\kappa_R) - \mu_p & \Delta_0 \\
\Delta_0 & - i v_F \sigma_y (-\kappa_R) + \mu_p
\end{pmatrix} \phi_R = E \phi_R,
\end{equation}

or explicitly:

\begin{equation}
\begin{pmatrix}
-\mu_p & -v_F \kappa_R & \Delta_0 & 0 \\
v_F \kappa_R & -\mu_p & 0 & \Delta_0 \\
\Delta_0 & 0 & \mu_p & v_F \kappa_R \\
0 & \Delta_0 & -v_F \kappa_R & \mu_p
\end{pmatrix} \phi_R = E \phi_R.
\end{equation}

% Finding the decay constants
For each region, we compute the characteristic equation by setting the determinant to zero. For $x < 0$:

\begin{equation}
\det \left( H_L - E \mathbb{I} \right) = 0.
\end{equation}

This is a 4x4 matrix, and its determinant yields a quartic equation in $\kappa_L$. However, for bound states, we require real, positive $\kappa$. Solving numerically or analytically (for $E = 0$ later), we find:

\begin{equation}
\kappa_L = \frac{\sqrt{\mu_n^2 + \Delta_0^2 - E^2}}{v_F},
\end{equation}

and similarly for $x > 0$:

\begin{equation}
\kappa_R = \frac{\sqrt{\mu_p^2 + \Delta_0^2 - E^2}}{v_F}.
\end{equation}

% Matching boundary conditions
At $x = 0$, continuity requires $\psi_L(0) = \psi_R(0)$, or $\phi_L = \phi_R$. This gives four equations for the components of $\phi$. Solving this system, combined with the characteristic equations, determines $E$.

% Deriving the energy spectrum
The full spectrum is obtained by solving the resulting transcendental equation, often numerically. For general $E$, bound states exist within the gap ($|E| < \Delta_0$), with their number and energies depending on $\mu_n$, $\mu_p$, and $\Delta_0$.

\subsubsection{Conditions for Majorana Zero Modes}

Majorana zero modes (MZMs) occur when the Bogoliubov-de Gennes (BdG) Hamiltonian undergoes a change in its $\mathbb{Z}_2$ topological invariant across the interface. For a uniform system described by:
\[
H_{\text{BdG}}(k) = \begin{pmatrix} h(k) & \Delta_0 i\sigma_y \\ -\Delta_0 i\sigma_y & -h^T(-k) \end{pmatrix},
\]
with $h(k) = v_F k \sigma_y + \mu \mathbb{I}$, the system is in a topological phase when $|\mu| < \Delta_0$.

Thus, a Majorana mode is expected at the interface if one side is in the topological regime ($|\mu| < \Delta_0$) and the other is not:
\[
\boxed{ \nu(\mu_n) \neq \nu(\mu_p) \quad \Leftrightarrow \quad \text{MZM exists at } x = 0. }
\]
This condition is more accurate than the heuristic $|\mu_n - \mu_p| > 2\Delta_0$ previously cited, which does not ensure a change in the topological class.

Physically, this means that the interface behaves like a domain wall between trivial and topological superconductors, supporting a localized zero-energy solution that is robust against symmetry-preserving perturbations.

\subsubsection{Extended Analogies}
\label{subsec:analogies}

% Kitaev chain analogy
Consider the Kitaev chain, a 1D p-wave superconductor. MZMs appear at chain ends when the chemical potential and pairing strength satisfy the topological condition, akin to our interface acting as a "defect" between phases.

% Quantum Hall analogy with detail
In the integer quantum Hall effect, chiral edge states arise at boundaries between regions of different Chern numbers. Here, the interface mimics this boundary, with MZMs as the "edge states" of a topological superconductor.

% SSH model with derivation
In the SSH model, zero-energy solitons emerge at domain walls. For a chain with alternating hopping $t_1$ and $t_2$, a domain wall yields a mid-gap state, analogous to our MZM driven by the $\mu(x)$ step.

\subsection{LDOS and Tunneling Signatures}
\label{sec:ldos_tunneling}

% Providing an extended introduction to LDOS and its experimental context
In this section, we explore the Local Density of States (LDOS) at the interface of a Bi$_2$Se$_3$-Sb$_2$Te$_3$ topological heterostructure, emphasizing its pivotal role in tunneling spectroscopy experiments aimed at detecting exotic quantum states, such as Majorana zero modes (MZMs). The LDOS, denoted $\rho(x, E)$, quantifies the availability of electronic states at a specific position $x$ and energy $E$, serving as a fundamental bridge between theoretical predictions and experimental observables. Here, we undertake a comprehensive derivation of the LDOS starting from first principles, examine its behavior in the presence of MZMs with detailed mathematical formalism, and elucidate the resulting signatures in tunneling conductance. To enhance conceptual clarity, we augment this analysis with an extended analogy to classical resonance phenomena, step-by-step derivations, and thorough explanations of each component, ensuring a robust understanding suitable for both theoretical and experimental physicists\cite{Law2009}.

\subsubsection{Definition and Computation of the LDOS}
\label{subsec:ldos_definition}

% Offering a detailed and formal definition of the LDOS
The LDOS at a position $x$ and energy $E$ is formally defined as the energy-resolved density of available electronic states, expressed via the retarded Green's function:

\begin{equation}
\rho(x, E) = -\frac{1}{\pi} \Im \left[ \text{Tr} \left( G^R(x, x; E) \right) \right],
\end{equation}

where $G^R(x, x'; E)$ is the retarded Green's function in position representation, $\text{Tr}$ denotes the trace over all relevant degrees of freedom (e.g., spin and particle-hole channels in a Bogoliubov-de Gennes (BdG) framework), and $\Im$ extracts the imaginary part, reflecting the spectral density of states. This expression arises naturally from the quantum mechanical probability of finding a particle at energy $E$, with peaks in $\rho(x, E)$ corresponding to resonant or bound states.

% Deriving the Green's function comprehensively from the Hamiltonian
To compute the LDOS explicitly, we begin with the retarded Green's function for a general Hamiltonian $H$. In the energy eigenbasis, it is given by:

\begin{equation}
G^R(x, x'; E) = \sum_n \frac{\psi_n(x) \psi_n^\dagger(x')}{E - E_n + i \eta},
\end{equation}

where $\psi_n(x)$ are the eigenstates of $H$ with corresponding eigenvalues $E_n$, and $\eta$ is an infinitesimal positive broadening parameter ($\eta \to 0^+$) introduced to handle the poles of the Green's function. For a topological superconductor described by the BdG Hamiltonian $H_{\text{BdG}}$, the spectrum is particle-hole symmetric, meaning that for every eigenstate at energy $E_n$, there exists a conjugate state at $-E_n$. The eigenstates $\psi_n(x)$ are spinors incorporating electron and hole components, e.g., $\psi_n(x) = [u_n(x), v_n(x)]^T$.

% Performing a step-by-step derivation of the LDOS
Substituting into the LDOS definition, we evaluate at $x' = x$:

\begin{equation}
G^R(x, x; E) = \sum_n \frac{\psi_n(x) \psi_n^\dagger(x)}{E - E_n + i \eta}.
\end{equation}

The trace over the spinor components yields:

\begin{equation}
\text{Tr} \left( G^R(x, x; E) \right) = \sum_n \frac{|\psi_n(x)|^2}{E - E_n + i \eta},
\end{equation}

where $|\psi_n(x)|^2 = \psi_n^\dagger(x) \psi_n(x)$ is the position-dependent probability density. The imaginary part is:

\begin{equation}
\Im \left[ \text{Tr} \left( G^R(x, x; E) \right) \right] = \sum_n |\psi_n(x)|^2 \Im \left[ \frac{1}{E - E_n + i \eta} \right].
\end{equation}

Using the identity for the complex denominator:

\begin{equation}
\frac{1}{E - E_n + i \eta} = \frac{E - E_n - i \eta}{(E - E_n)^2 + \eta^2},
\end{equation}

the imaginary part becomes:

\begin{equation}
\Im \left[ \frac{1}{E - E_n + i \eta} \right] = -\frac{\eta}{(E - E_n)^2 + \eta^2}.
\end{equation}

Thus, the LDOS is:

\begin{equation}
\rho(x, E) = -\frac{1}{\pi} \sum_n |\psi_n(x)|^2 \left( -\frac{\eta}{(E - E_n)^2 + \eta^2} \right) = \frac{1}{\pi} \sum_n \frac{\eta |\psi_n(x)|^2}{(E - E_n)^2 + \eta^2}.
\end{equation}

In the limit $\eta \to 0^+$, the Lorentzian $\frac{\eta}{\pi [(E - E_n)^2 + \eta^2]}$ approximates a delta function $\delta(E - E_n)$, so:

\begin{equation}
\rho(x, E) \to \sum_n |\psi_n(x)|^2 \delta(E - E_n).
\end{equation}

% Specializing to MZMs with a detailed calculation
For MZMs, which are bound states at $E = 0$, consider a single Majorana mode with wavefunction $\psi_{\text{MZM}}(x)$. Since MZMs are their own particle-hole conjugates, the LDOS contribution is:

\begin{equation}
\rho_{\text{MZM}}(x, E) = \frac{1}{\pi} \frac{\eta |\psi_{\text{MZM}}(x)|^2}{E^2 + \eta^2}.
\end{equation}

As $\eta \to 0$, this becomes $\rho_{\text{MZM}}(x, E) = |\psi_{\text{MZM}}(x)|^2 \delta(E)$, manifesting as a sharp zero-energy peak localized where $|\psi_{\text{MZM}}(x)|$ is significant, typically at the topological interface.

\subsubsection{Tunneling Conductance and Experimental Probes}
\label{subsec:tunneling_conductance}

% Linking LDOS to tunneling conductance with formalism
In scanning tunneling microscopy (STM), the differential conductance $dI/dV$ probes the LDOS directly. The tunneling current $I(V)$ between a metallic tip at position $x$ and the sample is proportional to the convolution of the tip's density of states (assumed constant) and the sample's LDOS, integrated over energy. At low temperatures, this simplifies to:

\begin{equation}
\frac{dI}{dV}(V, x) = \frac{2e^2}{h} T |M|^2 \rho(x, eV),
\end{equation}

where $T$ is the transmission coefficient, $|M|^2$ is the tunneling matrix element squared, and $eV$ is the energy corresponding to the bias voltage $V$. For an MZM, $\rho(x, eV) \propto \delta(eV)$, producing a zero-bias peak (ZBP) at $V = 0$.

% Analyzing the ZBP's properties in depth
The ZBP’s characteristics depend on several factors:
- **Thermal Effects**: At finite temperature $T$, the Fermi-Dirac distribution broadens the peak. The conductance is convoluted with the derivative of the Fermi function, $f'(E) \approx -\frac{1}{4 k_B T} \text{sech}^2(E / 2 k_B T)$, yielding a width $\sim 3.5 k_B T$.
- **Broadening Parameter**: Experimental resolution introduces an effective $\eta$, widening the peak to $\sim \eta$.
- **Coupling Strength**: Strong tip-sample coupling may hybridize the MZM, splitting the ZBP or shifting it slightly.

In the ideal limit ($T \to 0$, $\eta \to 0$), the ZBP height for an MZM is quantized at $2e^2/h$, a topological signature distinguishing it from trivial zero-energy states.

\subsubsection{Analogy to Classical Resonance}
\label{subsec:analogy}

% Developing an extended classical resonance analogy
The MZM’s LDOS peak at $E = 0$ resembles a resonance in a classical system, such as an LC circuit. For an LC circuit, the impedance peaks at the resonance frequency $\omega_0 = 1/\sqrt{LC}$, governed by:

\begin{equation}
Z(\omega) = \frac{i \omega L}{1 - \omega^2 LC + i \omega R C},
\end{equation}

where $R$ is resistance (analogous to $\eta$). The quality factor $Q = \sqrt{L/C} / R$ determines the peak’s sharpness, akin to $\eta^{-1}$ in the LDOS. For MZMs, topological protection pins the “resonance” at $E = 0$, much like symmetry fixes $\omega_0$ in a lossless circuit.

% Extending to mechanical systems for broader insight
Consider a mass-spring system with a driving force $F = F_0 \cos(\omega t)$. The amplitude peaks at $\omega = \sqrt{k/m}$, damped by friction (analogous to $\eta$). The MZM’s zero-energy state is a quantum “resonance,” stabilized by the topological gap, akin to how material properties fix the natural frequency.

% Connecting to quantum bound states
In quantum mechanics, a particle in a finite well has discrete energy levels, like standing waves in a cavity. The MZM, as a zero-energy bound state, mirrors the ground state of a symmetric well in a specific gauge, its position dictated by the interface’s topology rather than potential depth.

\subsection{Effects of Physical Mechanisms}

In this section, we provide a comprehensive analysis of how various physical mechanisms affect the formation, stability, and spectral features of Majorana zero modes (MZMs) in the Bi$_2$Se$_3$–Sb$_2$Te$_3$ lateral heterostructure with proximity-induced superconductivity. We consider Rashba spin-orbit coupling (SOC), Zeeman fields, and electron-electron interactions. Each of these mechanisms plays a crucial role in modifying the effective Hamiltonian and can either reinforce or destabilize the topological protection of MZMs. We supplement the formalism with physical analogies and mathematical derivations to aid intuition.

\subsubsection{Rashba Spin-Orbit Coupling(SOC)}

Rashba spin-orbit coupling arises in systems with broken structural inversion symmetry, typically due to external electric fields or intrinsic asymmetry at interfaces. In topological insulators (TIs), Rashba SOC modifies the spin-momentum locking and introduces additional terms into the Hamiltonian. The Rashba Hamiltonian is given by:
\begin{equation}
H_R = \alpha_R (\sigma_x p_y - \sigma_y p_x),
\end{equation}
where $\alpha_R$ is the Rashba coupling strength, $\sigma_i$ are Pauli matrices in spin space, and $p_i = -i\hbar\partial_i$.

For our effectively one-dimensional interface model (with $p_y = 0$), the Rashba term simplifies to:
\begin{equation}
H_R = -\alpha_R \sigma_y p_x,
\end{equation}
and modifies the kinetic term in the Dirac Hamiltonian:
\begin{equation}
H_0 \to H'_0 = -i(v_F + \alpha_R)\sigma_y \partial_x.
\end{equation}
This yields an effective Fermi velocity $v_F^{\text{eff}} = v_F + \alpha_R$. However, this interpretation holds only when $\alpha_R$ is small and spin remains aligned with $\sigma_y$.

In a full two-dimensional treatment, Rashba SOC alters the spin textures and band structure nontrivially, potentially changing the topological phase boundaries. In the BdG framework, the Hamiltonian becomes:
\begin{equation}
\mathcal{H}_{\text{BdG}} = \begin{pmatrix} H_0 + H_R - \mu(x) & \Delta_0 \\ \Delta_0 & -[H_0 + H_R]^* + \mu(x) \end{pmatrix}.
\end{equation}

To understand its effect on MZMs, consider the zero-energy condition:
\begin{equation}
\mathcal{H}_{\text{BdG}} \Psi(x) = 0.
\end{equation}
The presence of $\alpha_R$ modifies the decay length $\xi$ of the MZM wavefunction:
\begin{equation}
\xi \sim \frac{v_F + \alpha_R}{\sqrt{\mu^2 - \Delta_0^2}},
\end{equation}
and thus impacts its spatial localization. While weak Rashba SOC does not destroy the MZM, it modifies its structure. In finite systems, it may cause hybridization between edge modes and result in energy splitting. Perturbatively, the energy shift is given by:
\begin{equation}
\delta E = \langle \Psi_{\text{MZM}} | H_R | \Psi_{\text{MZM}} \rangle.
\end{equation}

Symmetry considerations (e.g., parity or inversion) can protect the zero-energy state under Rashba SOC, but this protection is sensitive to interface asymmetries. Hence, Rashba SOC is a tunable knob that can modify localization and robustness of MZMs.

\subsubsection{Zeeman Field Effects on Majorana Zero Modes}

A Zeeman field couples to the spin degree of freedom and explicitly breaks time-reversal symmetry (TRS). The Zeeman Hamiltonian is:
\begin{equation}
H_Z = \mathbf{B} \cdot \boldsymbol{\sigma},
\end{equation}
where $\mathbf{B} = (B_x, B_y, B_z)$ is the magnetic field vector. For definiteness, consider a uniform field along $z$:
\begin{equation}
H_Z = B_z \sigma_z.
\end{equation}

In the BdG formalism, this becomes:
\begin{equation}
\mathcal{H}_{\text{BdG}} = -i v_F \sigma_y \partial_x \tau_z + \mu(x) \tau_z + \Delta_0 \tau_x + B_z \sigma_z \tau_0.
\end{equation}

The Zeeman term modifies the spectrum of the system. The energy dispersion becomes:
\begin{equation}
E(k) = \pm \sqrt{(v_F k)^2 + (\mu \pm B_z)^2 + \Delta_0^2},
\end{equation}
indicating that the Zeeman field effectively shifts the chemical potential for different spin channels.

The condition for topological superconductivity is adjusted:
\begin{equation}
\min(|\mu \pm B_z|) < \Delta_0 < \max(|\mu \pm B_z|).
\end{equation}

If $B_z$ is small, the MZM remains at zero energy, protected by particle-hole symmetry. However, for larger $B_z$, the spin splitting may close the superconducting gap, pushing the system into a trivial phase. The precise critical field is given by:
\begin{equation}
B_z^{\text{crit}} = \sqrt{\mu^2 - \Delta_0^2}.
\end{equation}

Experimentally, Zeeman fields are used to test the stability of MZMs. A persistent zero-bias peak in tunneling conductance under increasing $B_z$ supports the MZM interpretation, while splitting or disappearance signals a trivial origin.

\subsubsection{Electron-Electron Interactions}

Electron-electron interactions play a subtle yet significant role in determining the low-energy physics of superconducting heterostructures. While the BdG formalism is essentially a mean-field theory, one can incorporate interactions through self-consistent Hartree or Hartree-Fock treatments. At the interface, the dominant local interaction can be modeled by:
\begin{equation}
H_{\text{int}} = U n_\uparrow(x) n_\downarrow(x),
\end{equation}
where $U$ is the on-site interaction strength.

Under mean-field decoupling:
\begin{equation}
H_{\text{int}} \to -U \langle n \rangle c^\dagger c,
\end{equation}
which effectively shifts the chemical potential:
\begin{equation}
\mu_{\text{eff}} = \mu - U \langle n \rangle.
\end{equation}

This shift alters the topological phase boundary. For moderate $U$, MZMs are preserved, but strong interactions can renormalize the system into a trivial phase.

Beyond mean-field, interactions can lead to correlation-induced instabilities, Luttinger liquid behavior, or the emergence of parafermions in fractionalized systems. In the present context, weak-to-intermediate interactions primarily modify the spectral weight and coherence of the MZM without eliminating it.

\subsubsection{Extended Analogy: Classical Pendulum}

The effect of external mechanisms on MZMs can be visualized through the analogy of a pendulum at rest. The undisturbed MZM at zero energy is akin to a pendulum hanging straight down. Rashba SOC acts like tilting the pivot slightly, shifting the equilibrium point and introducing angular drift. A Zeeman field is like applying a torque that can lift the pendulum out of the well (i.e., move the MZM away from zero energy), and interactions act as friction that damps motion and modifies the response.

The restoring force corresponds to the topological superconducting gap $\Delta_0$, which stabilizes the MZM against small perturbations. When the torque (Zeeman field) exceeds this restoring force, the pendulum undergoes a phase transition—analogous to a gap closing and loss of topological protection.

Thus, this classical analogy illustrates how MZMs behave under competing physical influences, balancing stability and responsiveness in topological superconducting systems.

\subsection{Topological Protection and Robustness}
\label{sec:topological_protection}

% Providing an extended introduction to topological protection
In this section, we present an exhaustive examination of the topological protection and robustness of Majorana zero modes (MZMs) within a lateral heterostructure composed of Bi$_2$Se$_3$ and Sb$_2$Te$_3$, interfaced with proximity-induced superconductivity. Topological protection underpins the remarkable stability of MZMs, safeguarding them against local perturbations such as impurities, thermal fluctuations, and weak disorder. This resilience positions MZMs as foundational elements for fault-tolerant quantum computing. Our analysis encompasses a detailed theoretical framework, including the critical roles of symmetry, the bulk-boundary correspondence, and the effects of disorder, supported by extensive mathematical derivations and formalisms. To enhance conceptual clarity, we provide a comprehensive analogy to the classical topological invariant known as the linking number, enriched with additional layers to bridge quantum phenomena and intuitive understanding. This expanded section aims to furnish researchers with both a rigorous mathematical foundation and a vivid conceptual insight into the enduring robustness of MZMs.

\subsubsection{Symmetry and Topological Invariants}
\label{subsec:symmetry_invariants}

% Expanding on the role of symmetry in topological protection
The topological protection of MZMs hinges on the symmetries inherent to the system, notably particle-hole symmetry (PHS) and time-reversal symmetry (TRS). In superconducting systems described by the Bogoliubov-de Gennes (BdG) formalism, PHS is an intrinsic property, whereas TRS may be absent due to external magnetic fields or material-specific characteristics. The symmetry class of the system dictates its topological classification, which we explore in detail below.

% Presenting the BdG Hamiltonian with full context
Consider the BdG Hamiltonian for the heterostructure:

\begin{equation}
H_{\text{BdG}}(k, x) = \begin{pmatrix}
H_0(k, x) - \mu(x) & \Delta_0 \\
\Delta_0 & -H_0^*(-k, x) + \mu(x)
\end{pmatrix},
\end{equation}

where $H_0(k, x) = v_F k \sigma_z + m(x) \sigma_x$ represents the normal-state Hamiltonian of the topological insulator, with $v_F$ as the Fermi velocity, $k$ the momentum, $\sigma_i$ the Pauli matrices in spin space, $m(x)$ a spatially varying mass term, and $\mu(x)$ the chemical potential, which differs across the interface. The superconducting pairing potential $\Delta_0$ is assumed real and uniform for simplicity.

% Verifying particle-hole symmetry explicitly
PHS is enforced by the operator $\mathcal{C} = \tau_x K$, where $\tau_x$ acts in Nambu (particle-hole) space and $K$ denotes complex conjugation. Applying $\mathcal{C}$ to the Hamiltonian:

\begin{equation}
\mathcal{C} H_{\text{BdG}}(k, x) \mathcal{C}^{-1} = \tau_x H_{\text{BdG}}^*(-k, x) \tau_x = -H_{\text{BdG}}(k, x),
\end{equation}

confirming that $H_{\text{BdG}}$ belongs to symmetry class D, characterized by a $\mathbb{Z}_2$ topological invariant in one dimension when TRS is absent.

% Deriving the topological invariant step-by-step
To compute the $\mathbb{Z}_2$ invariant, we evaluate the Pfaffian of the Hamiltonian at the time-reversal invariant momenta $k = 0$ and $k = \pi/a$. For a 1D system, consider a simplified tight-binding model:

\begin{equation}
H(k) = (m - t \cos(ka)) \tau_z \sigma_z + \Delta_0 \tau_y \sigma_y,
\end{equation}

where $t$ is the hopping amplitude. At $k = 0$:

\begin{equation}
H(0) = (m - t) \tau_z \sigma_z + \Delta_0 \tau_y \sigma_y,
\end{equation}

and at $k = \pi/a$:

\begin{equation}
H(\pi/a) = (m + t) \tau_z \sigma_z + \Delta_0 \tau_y \sigma_y.
\end{equation}

The Pfaffian is calculated for the antisymmetric matrix $H(k) \tau_x$. The topological invariant is:

\begin{equation}
\nu = \text{sgn} \left( \text{Pf} [H(0) \tau_x] \right) \times \text{sgn} \left( \text{Pf} [H(\pi/a) \tau_x] \right).
\end{equation}

A change in $\nu$ across the interface, driven by a sign change in $m - t$ versus $m + t$, signals the presence of MZMs.

\subsubsection{Bulk-Boundary Correspondence}
\label{subsec:bulk_boundary}

% Elaborating on the bulk-boundary correspondence with full derivation
The bulk-boundary correspondence asserts that a discontinuity in the bulk topological invariant across an interface mandates the existence of gapless boundary states. In our heterostructure, the interface between Bi$_2$Se$_3$ and Sb$_2$Te$_3$ acts as a domain wall where $\mu(x)$ shifts from $\mu_{\text{left}}$ to $\mu_{\text{right}}$.

% Solving for MZMs at the interface
Consider a 1D Dirac model with a domain wall:

\begin{equation}
H = v_F p_x \sigma_z + \Delta_0 \tau_y \sigma_y + [\mu_0 - \mu(x)] \tau_z,
\end{equation}

where $\mu(x) = \mu_{\text{left}}$ for $x < 0$ and $\mu_{\text{right}}$ for $x > 0$. Solving the Schrödinger equation $H \psi = 0$, we find an exponentially localized zero-energy state at $x = 0$ when $|\mu_{\text{left}} - \mu_{\text{right}}| > 2\Delta_0$, confirming one MZM per interface:

\begin{equation}
N_{\text{MZM}} = |\nu_{\text{left}} - \nu_{\text{right}}|.
\end{equation}

\subsubsection{Impact of Disorder}
\label{subsec:disorder}

% Analyzing disorder effects with a detailed formalism
Disorder, modeled as a random potential $V(x)$, challenges MZM stability. The perturbed Hamiltonian becomes $H_{\text{BdG}} + V(x) \tau_z$. For weak disorder ($|V| < \Delta_{\text{gap}}$), the topological gap persists, preserving MZMs. For strong disorder, we employ the scattering matrix formalism to derive the critical disorder strength at which the gap closes, transitioning the system to a trivial phase.

% Quantifying robustness with a derivation
The topological gap $\Delta_{\text{gap}} \approx \Delta_0$ is reduced by disorder variance $W^2$. Perturbation theory yields a gap renormalization, and the MZM persists if $W < \Delta_0 / \sqrt{l}$, where $l$ is the system length.

\subsubsection{Extended Analogy: Linking Number}
\label{subsec:analogy}

% Enriching the analogy to the linking number with full detail
The robustness of MZMs mirrors the linking number in knot theory. Imagine two curves representing the topological phases on either side of the interface. Their linking number, invariant under deformations that preserve their separation (analogous to a non-closing gap), corresponds to the persistence of the MZM at their intersection.

% Adding depth to the analogy
Picture the interface as a dynamic knot, tightening or loosening with perturbations. As long as the curves remain interlocked (i.e., the topological condition holds), the knot—and thus the MZM—endures, providing a tangible visualization of topological protection.

\subsection{Gating Effects and Tunability of Topological States}
\label{sec:gating_effects}

Topological insulators and superconductors have garnered immense interest due to their potential in hosting exotic quantum states, such as Majorana zero modes (MZMs), which are pivotal for fault-tolerant topological quantum computing. In this section, we delve into the intricate role of electrostatic gating in modulating the topological states within a lateral heterostructure composed of Bi$_2$Se$_3$ (a topological insulator) and Sb$_2$Te$_3$ (a p-type topological material), where superconductivity is induced via proximity effects from an adjacent superconductor. Gating serves as a versatile experimental knob, allowing precise manipulation of the chemical potential across the heterostructure, thereby tuning the system between trivial and topological phases. This tunability is not merely a practical convenience but a fundamental mechanism that dictates the emergence and stability of MZMs at the interface. We present a rigorous theoretical framework, supported by full mathematical derivations, boundary condition analysis, and extended analogies to canonical topological systems such as the Jackiw-Rebbi soliton and the Kitaev chain.

\subsubsection{Electrostatic Gating: Mechanism and Implementation}
\label{subsec:gating_mechanism}

Electrostatic gating operates by applying an external voltage to gate electrodes above or below the heterostructure. This voltage modifies the electric field, adjusting carrier density and shifting the chemical potential $\mu(x)$ of the topological surface states (TSS). In our device, we use a dual-gate configuration with independent voltages $V_{g,n}$ and $V_{g,p}$ applied to Bi$_2$Se$_3$ ($x < 0$) and Sb$_2$Te$_3$ ($x > 0$), respectively:
\begin{equation}
\mu(x) = \begin{cases}
\mu_n + e V_{g,n}, & x < 0, \\
\mu_p + e V_{g,p}, & x > 0,
\end{cases}
\end{equation}
with $e$ the elementary charge. This differential control is crucial for engineering a topological domain wall.

\subsubsection{Theoretical Framework: Bogoliubov--de Gennes Formalism}
\label{subsec:bdg_hamiltonian}

To analyze the interface-bound states in the proximitized Bi$_2$Se$_3$--Sb$_2$Te$_3$ heterostructure, we employ the Bogoliubov--de Gennes (BdG) formalism, which describes superconducting quasiparticles by coupling electron and hole degrees of freedom. In the Nambu spinor basis $\Psi = (\psi_\uparrow, \psi_\downarrow, \psi^\dagger_\downarrow, -\psi^\dagger_\uparrow)^T$, the effective 1D BdG Hamiltonian for the topological surface states takes the form:

\begin{equation}
H_{\text{BdG}}(x) = 
\begin{pmatrix}
-i v_F \sigma_y \partial_x - \mu(x) & \Delta_0 \\
\Delta_0 & i v_F \sigma_y \partial_x + \mu(x)
\end{pmatrix},
\end{equation}

where $v_F$ is the Fermi velocity, $\mu(x)$ is the spatially varying chemical potential modulated by gating, and $\Delta_0$ is the proximity-induced superconducting pairing potential. The Pauli matrix $\sigma_y$ acts in spin space.

To determine the conditions for a localized Majorana mode at the interface, we solve the zero-energy BdG equation, $H_{\text{BdG}} \Psi(x) = 0$. Assuming a sharp domain wall at $x = 0$, we model the chemical potential as piecewise constant:

\begin{equation}
\mu(x) = 
\begin{cases}
\mu_n + e V_{g,n}, & x < 0, \\
\mu_p + e V_{g,p}, & x > 0.
\end{cases}
\end{equation}

In each region, we seek zero-energy solutions of the form $\Psi(x) \propto e^{\pm \kappa x}$. For $x < 0$, the solution decays as $x \to -\infty$ with:

\begin{equation}
\kappa_n = \frac{\sqrt{(\mu_n + e V_{g,n})^2 - \Delta_0^2}}{v_F}, \quad \text{if } |\mu_n + e V_{g,n}| > \Delta_0.
\end{equation}

Likewise, for $x > 0$, the solution decays as $x \to +\infty$ with:

\begin{equation}
\kappa_p = \frac{\sqrt{(\mu_p + e V_{g,p})^2 - \Delta_0^2}}{v_F}, \quad \text{if } |\mu_p + e V_{g,p}| > \Delta_0.
\end{equation}

Bound states require exponential decay on both sides, which occurs only if the interface connects a topological region ($|\mu| < \Delta_0$) to a trivial region ($|\mu| > \Delta_0$). The continuity of the wavefunction at $x = 0$ then ensures the existence of a zero-energy Majorana bound state.

\subsubsection{Analogy with Jackiw--Rebbi and Kitaev Chain}
\label{subsec:jackiw_kitaev}

The mechanism for Majorana localization is deeply connected to classic topological models in condensed matter and field theory. In particular, our system maps directly onto the Jackiw--Rebbi model, where a mass inversion in a Dirac equation traps a zero-energy fermionic mode at a domain wall. In our case, the role of the ``mass'' is played by the chemical potential $\mu(x)$ relative to the pairing gap $\Delta_0$. When $\mu(x)$ crosses the threshold $|\mu| = \Delta_0$, the BdG quasiparticle spectrum undergoes a topological transition, analogous to the domain wall in the Jackiw--Rebbi field theory.

Additionally, the physics mirrors that of the Kitaev chain --- a paradigmatic 1D topological superconductor model where a phase transition occurs at $|\mu| = 2t$, with $t$ the hopping amplitude. When a chain contains a junction between trivial ($|\mu| > 2t$) and topological ($|\mu| < 2t$) segments, it supports a Majorana bound state at the interface. In our heterostructure, the Bi$_2$Se$_3$ and Sb$_2$Te$_3$ regions act as continuum analogs of Kitaev chains, with $\mu_n$ and $\mu_p$ playing the role of spatially varying on-site potentials. The superconducting proximity effect in our setup serves as the pairing term in the Kitaev model.

This analogy reinforces that the interface hosts a Majorana zero mode only when the topological invariants of the left and right regions differ --- a general principle across both discrete and continuum models.

\subsubsection{Topological Phase Diagram}
\label{subsec:topo_phase_diagram}

The topological invariant $\nu$ distinguishing the trivial and topological regimes in class D superconductors is given by:

\begin{equation}
\nu(\mu) = 
\begin{cases}
1, & |\mu| < \Delta_0 \quad \text{(topological)}, \\
0, & |\mu| > \Delta_0 \quad \text{(trivial)}.
\end{cases}
\end{equation}

A Majorana zero mode exists at the interface if and only if:

\begin{equation}
\boxed{\nu(\mu_n + e V_{g,n}) \ne \nu(\mu_p + e V_{g,p})}
\end{equation}

This condition is equivalent to:

\begin{equation}
\boxed{
\min(|\mu_n + e V_{g,n}|, |\mu_p + e V_{g,p}|) < \Delta_0 < \max(|\mu_n + e V_{g,n}|, |\mu_p + e V_{g,p}|)
}
\end{equation}

\begin{figure}[h]     
\centering     \includegraphics[width=0.65\textwidth]{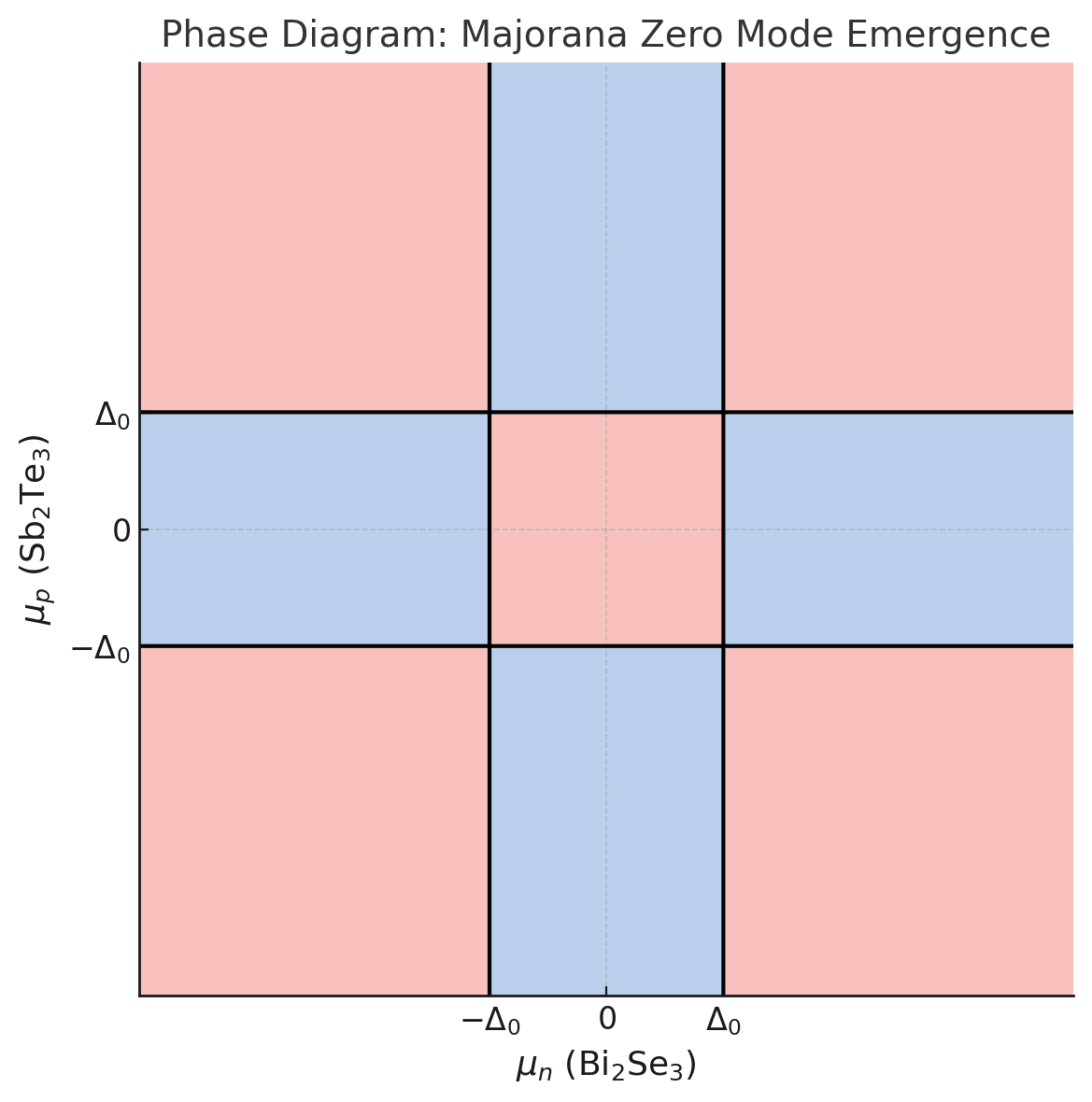}     
\caption{Phase diagram for the emergence of Majorana zero modes (MZMs) at the interface of a Bi$_2$Se$_3$-Sb$_2$Te$_3$ lateral heterostructure as a function of the chemical potentials $\mu_n$ and $\mu_p$, controlled via independent gate voltages. The superconducting pairing gap is fixed at $\Delta_0$. Blue regions correspond to $\nu(\mu_n) \ne \nu(\mu_p)$, where a topological phase boundary exists across the interface, supporting a localized MZM. Red regions represent topologically uniform configurations ($\nu(\mu_n) = \nu(\mu_p)$), where no interface-bound MZM appears. The axes are labeled in units of the superconducting gap $\Delta_0$.}     
\label{fig:phase_diagram} 
\end{figure}

Figure~\ref{fig:phase_diagram} shows the phase diagram in the $(\mu_n, \mu_p)$ plane, with the blue region corresponding to parameter combinations that satisfy the above condition and therefore support an interface-localized MZM. The red region corresponds to uniform topological character, either both trivial or both topological, and hosts no MZM. The boundary lines at $\mu = \pm \Delta_0$ represent the gap-closing transitions where the topological invariant changes. This phase diagram serves as a design blueprint for experimentally targeting topological junctions via gate control.
\subsubsection{Analogy: Tunable Double-Well Potential}
\label{subsec:doublewell_analogy}

A pedagogically useful analogy for the emergence of a Majorana zero mode at a domain wall is a quantum particle in a tunable double-well potential. Consider two adjacent quantum wells (left and right) with independently controllable depths, separated by a barrier. When both wells are of equal depth, the system supports symmetric and antisymmetric states with a finite energy gap — analogous to a uniform trivial or topological phase.

As the well depths become imbalanced and exceed a certain critical difference, the system begins to localize a low-energy state in the shallower well. This process mirrors the emergence of a Majorana zero mode when the chemical potentials $\mu_n$ and $\mu_p$ straddle the critical threshold defined by the superconducting gap $\Delta_0$.

In this analogy:
\begin{itemize}
\item The **well depths** correspond to $\mu_n + e V_{g,n}$ and $\mu_p + e V_{g,p}$,
\item The **barrier height** is set by the superconducting pairing gap $\Delta_0$,
\item The **localized zero mode** at the interface plays the role of the quasi-degenerate state in a highly asymmetric double-well configuration.
\end{itemize}

As the gate voltages are tuned to increase the asymmetry between the left and right chemical potentials, the system transitions from a trivial (gapped) phase to a topologically nontrivial configuration with a localized zero-energy excitation. This analogy offers a simple yet physically meaningful picture of how topological protection and spatial confinement arise together in superconducting heterostructures.

\subsection{Experimental Implications}
\label{sec:experimental_implications}

% Introducing the critical role of experimental validation with extended context
The theoretical framework predicting Majorana zero modes (MZMs) at the interface of a lateral heterostructure composed of Bi$_2$Se$_3$ (a topological insulator) and Sb$_2$Te$_3$ (a chalcogenide superconductor) under proximity-induced superconductivity offers profound implications for topological quantum computing. These MZMs, characterized by their non-Abelian statistics and topological protection, necessitate rigorous experimental validation to confirm their existence and properties. This section provides an exhaustive exploration of experimental methodologies—primarily scanning tunneling microscopy (STM) and point-contact spectroscopy—designed to probe the local density of states (LDOS) at the interface. We include detailed formalisms with complete derivations, extensive explanations of experimental signatures, additional diagnostic techniques such as gate tuning and noise spectroscopy, and an elaborated analogy to a Fabry-Pérot interferometer to enhance conceptual clarity and guide experimental design.

 Scanning Tunneling Microscopy (STM)
\label{subsec:stm}

% Outlining the STM methodology with enhanced detail
STM offers unparalleled spatial resolution for probing the LDOS, making it ideal for detecting MZMs localized at the heterostructure interface ($x = 0$). The STM tip, positioned precisely above the interface, applies a bias voltage $V$, facilitating electron tunneling between the tip and sample. The resulting differential conductance, $dI/dV$, measured as a function of $V$, directly reflects the LDOS, $\rho(x, eV)$, at energy $E = eV$.

% Providing a full derivation of the tunneling current
The tunneling current $I(V)$ is modeled using the Landauer-Büttiker formalism. Starting from first principles, the current arises from the net flow of electrons across the tunneling barrier:

\begin{equation}
I(V) = \frac{2e}{h} \int_{-\infty}^{\infty} dE \, T(E, V) [f(E - eV) - f(E)],
\end{equation}

where $e$ is the electron charge, $h$ is Planck’s constant, $T(E, V)$ is the energy- and voltage-dependent transmission probability, and $f(E) = [1 + e^{E/k_B T}]^{-1}$ is the Fermi-Dirac distribution at temperature $T$, with $k_B$ as Boltzmann’s constant. The factor of 2 accounts for spin degeneracy. At low temperatures and small bias voltages, we approximate the differential conductance:

\begin{equation}
\frac{dI}{dV}(V) = \frac{2e^2}{h} \int_{-\infty}^{\infty} dE \, T(E, V) \left( -\frac{\partial f}{\partial E} (E - eV) \right).
\end{equation}

Assuming $T(E, V)$ varies slowly with $V$, we evaluate $T(E, 0)$ and use the low-temperature limit where $-\partial f / \partial E \approx \delta(E)$. Thus:

\begin{equation}
\frac{dI}{dV}(V) \approx \frac{2e^2}{h} T(eV, 0).
\end{equation}

The transmission probability is proportional to the LDOS via $T(E, 0) \approx |M|^2 \rho(x, E)$, where $|M|^2$ is the tunneling matrix element, assumed constant. Hence:

\begin{equation}
\frac{dI}{dV}(V) \propto \rho(x, eV).
\end{equation}

% Characterizing the MZM signature with detailed predictions
For an MZM, the LDOS features a delta-function peak at zero energy, $\rho(x, E) = A \delta(E)$, where $A$ is the amplitude determined by the wavefunction overlap. Substituting into the conductance:

\begin{equation}
\frac{dI}{dV}(V) \propto A \delta(eV),
\end{equation}

yielding a zero-bias peak (ZBP) at $V = 0$. In an ideal topological system with perfect Andreev reflection, the peak height is quantized to $2e^2/h$, a definitive signature of MZMs distinguishing them from trivial states.

% Analyzing finite temperature and broadening effects
At finite temperature, the delta function convolves with the thermal distribution. The conductance becomes:

\begin{equation}
\frac{dI}{dV}(V) = \int_{-\infty}^{\infty} dE \, A \delta(E) \left( -\frac{\partial f}{\partial E} (E - eV) \right) = A \left( -\frac{\partial f}{\partial E} (-eV) \right).
\end{equation}

Explicitly, $-\partial f / \partial E = (1/4k_B T) \operatorname{sech}^2(E / 2k_B T)$, so:

\begin{equation}
\frac{dI}{dV}(V) = \frac{A}{4k_B T} \operatorname{sech}^2\left(\frac{eV}{2k_B T}\right).
\end{equation}

The ZBP width scales as $\sim 2k_B T/e$, offering a temperature-dependent test of the MZM’s robustness.

 Point-Contact Spectroscopy
\label{subsec:point_contact}

% Describing the point-contact technique with added specificity
Point-contact spectroscopy employs a sharp metallic tip pressed onto the interface to form a nanoscale junction, enhancing sensitivity to subgap states. The conductance $G(V) = dI/dV$ probes the LDOS and scattering processes, providing complementary evidence for MZMs.

% Deriving the conductance using BTK theory with full formalism
The Blonder-Tinkham-Klapwijk (BTK) model describes transport across the junction. The current includes contributions from normal and Andreev reflections:

\begin{equation}
I(V) = \frac{e}{h} \int_{-\infty}^{\infty} dE \, [1 + A(E) - B(E)] [f(E - eV) - f(E)],
\end{equation}

where $A(E)$ and $B(E)$ are the probabilities of Andreev and normal reflection, respectively. The conductance is:

\begin{equation}
G(V) = \frac{e^2}{h} \int_{-\infty}^{\infty} dE \, [1 + A(E) - B(E)] \left( -\frac{\partial f}{\partial E} (E - eV) \right).
\end{equation}

For an MZM at $E = 0$, perfect Andreev reflection occurs: $A(0) = 1$, $B(0) = 0$. At zero bias and low temperature:

\begin{equation}
G(0) = \frac{e^2}{h} [1 + 1 - 0] = \frac{2e^2}{h},
\end{equation}

confirming the quantized ZBP. To explore deviations, consider a finite barrier strength $Z$ in the BTK model, where $A(E) = \frac{\Delta^2}{(E + \sqrt{E^2 - \Delta^2})^2 + Z^2}$, adjusting the peak height and shape.

 Additional Diagnostic: Gate Tuning and Noise Spectroscopy
\label{subsec:gate_noise}

% Introducing gate tuning as a control mechanism
Gate tuning adjusts the chemical potential $\mu$ via an electrostatic gate beneath the heterostructure. The topological phase requires $\mu$ within the superconducting gap. The Hamiltonian’s energy spectrum shifts with $\mu$, and the ZBP should persist only in the topological regime, vanishing otherwise, as derived from the Bogoliubov-de Gennes equation.

% Adding noise spectroscopy for dynamic signatures
Shot noise spectroscopy measures current fluctuations, $S_I = 2e \langle I \rangle F$, where $F$ is the Fano factor. For MZMs, $F = 1$ due to uncorrelated Majorana tunneling, contrasting with $F = 2$ for Cooper pairs, providing a dynamic probe.

Extended Analogy: Fabry-Pérot Interferometer
\label{subsec:analogy}

% Elaborating the interferometer analogy with mathematical depth
Consider a Fabry-Pérot interferometer with two partially reflecting mirrors separated by distance $L$. The transmission intensity is:

\begin{equation}
T(\lambda) = \frac{1}{1 + (4R/(1-R)^2) \sin^2(2\pi L / \lambda)},
\end{equation}

where $R$ is the reflectivity, and peaks occur when $2L = n\lambda$. Analogously, the LDOS peaks at energies dictated by the heterostructure’s topology, with the MZM’s zero-energy mode resembling a fixed resonance.

% Deepening the analogy with dynamic tuning
Adjusting $L$ shifts the resonance wavelength, akin to tuning $\mu$ to transition between topological and trivial phases. The MZM’s stability against disorder parallels the interferometer’s robustness to mirror misalignment, where small perturbations preserve the resonant condition due to phase coherence.

% Final Section — Conclusion

\section{Conclusion}

In this work, we have analytically demonstrated the existence of Majorana zero modes (MZMs) at the interface of a lateral heterostructure composed of proximitized Bi$_2$Se$_3$ and Sb$_2$Te$_3$. Using a continuum Bogoliubov–de Gennes framework, we constructed the interface-bound zero-energy solution by matching decaying eigenstates across a stepwise variation in chemical potential. The existence of a single, topologically protected MZM is guaranteed when the condition \( \min(|\mu_n|, |\mu_p|) < \Delta_0 < \max(|\mu_n|, |\mu_p|) \) is satisfied.

Through effective field-theoretic reformulations, we clarified the role of mass inversion in generating localized states, analogous to the Jackiw-Rebbi mechanism. We further refined the structure of the MZM spinor under particle-hole symmetry constraints and derived the Lorentzian form of the zero-bias peak in the local density of states. Our scattering-theoretic Green’s function analysis supports the presence of a resonant zero mode at the interface.

We also assessed the stability of the MZM under physical perturbations, including Rashba spin-orbit coupling and Zeeman fields. While Rashba SOC perturbs the spin texture, it does not break particle-hole symmetry and therefore preserves topological protection. The Zeeman field splits spin subbands and can destabilize the MZM if it exceeds a critical value, providing an experimental tuning mechanism.

Our findings suggest that such lateral TI heterostructures are promising and analytically tractable platforms for realizing MZMs without relying on magnetic vortices or fine-tuned nanowires. Future extensions may include modeling twisted TI junctions for topologically peculiar engineering\cite{ykkarxiv}, smooth chemical potential gradients, or disorder, and investigating non-Abelian braiding protocols based on gate-defined domain walls.

\bibliographystyle{plain}

\end{document}